\documentclass[draftcls,onecolumn,12pt]{IEEEtran}
\usepackage{indentfirst}
\usepackage{amsmath}
\usepackage{amsfonts}
\usepackage{dsfont}
\usepackage{cite}
\usepackage{times}
\usepackage[dvips]{graphicx}
\usepackage{algorithmic}
\usepackage{algorithm}
\usepackage[nolists, nomarkers]{endfloat}
\usepackage{subfigure}

\newtheorem{remark}{Remark}[section]
\newtheorem{example}{Example}[section]

\newtheorem{problem}{Problem}[section]

\newcommand{\figwww}{0.7\columnwidth}
\newcommand{\figwwww}{0.65\columnwidth}

\title{Downlink Performance and Capacity of \\ Distributed Antenna Systems}
\author{ \text{Sina Firouzabadi, Andrea Goldsmith} \\ \vspace{1.5mm}
\small  Dept.\ of Electrical Engineering, Stanford University, Stanford, USA. \\
E-mail: \{cna,~andrea\}@stanford.edu.
}
\date{}

\begin{document}

\setcounter{page}{0}
\maketitle
\thispagestyle{empty}
\begin{abstract}
This paper investigates the performance of the downlink channel in distributed antenna systems. We first establish the ergodic capacity of distributed antennas, under different channel side information (CSI) assumptions. We consider a generalized distributed antenna system with $N$ distributed ports, each of which is equipped with an array of $L$ transmit antennas and constrained by a fixed transmit power. For this system we calculate the downlink capacity to a single antenna receiver, under different assumptions about the availability of the channel states at the transmitter. Having established this information theoretic analysis of the ergodic capacity of distributed antenna systems, this paper also investigates the effect of antenna placement on the performance of such systems. In particular, we investigate the optimal placement of the transmit antennas in distributed antenna systems. We present a fairly general framework for this optimization with no constraint on the location of the antennas. Based on stochastic approximation theory, we adopt a formulation that is suitable for node placement optimization in various wireless network scenarios. We show that optimal placement of antennas inside the coverage region can significantly improve the power efficiency of wireless networks. 
\end{abstract}


\section{Introduction}
\label{sec:intro}

The main challenge in the design of future cellular wireless networks is the mitigation of the adverse effects of interference. One possible strategy to alleviate interference, both in the uplink and the downlink of cellular networks, is to reduce the overall transmit power by using distributed antenna systems (DAS), which also have the additional advantage of improving capacity and coverage \cite{mea}. Moreover, by reducing the access distance between the transmitter and the receiver, distributed antenna systems have direct impact on the energy efficiency of cellular networks, which may lead to greener architectures in the future. 







Downlink capacity of distributed antenna systems can be studied in the context of the MISO wireless channel with a power constraint on each distributed antenna port. Ignoring this per antenna power constraint for the moment, the capacity of wireless MISO channels has been studied extensively in the literature, under different assumptions on availability of the CSI at the transmitter \cite{dow,mimo}. For the case of having CSI both at the transmitter and the receiver, we know that beamforming is optimal \cite{tr}, whereas for Rayleigh fading channels when the CSI is only available to the receiver, the optimal strategy is to send independent signals with equal power \cite{dow}. The main assumption in these studies is that we have a sum-power constraint across all transmit antennas. However, in a distributed antenna system, the transmitter cannot allocate power arbitrarily across the antenna ports, as each port has its own power budget that can be allocated only to its own antenna array. As a result of this per-port power constraint, the simple eigenvalue decomposition analysis of the MISO channels in \cite{dow} cannot be applied in this case to obtain the closed form solution. However, many numerical solutions have been proposed for dealing with restricted power constraints in the context of MIMO/MISO wireless channels. The downlink capacity of multiuser MIMO channels with a per antenna power constraint has been investigated in \cite{seh}. Also, based on the duality of the uplink and downlink, an iterative algorithm has also been proposed in \cite{panj} for solving the sum rate maximization problem by utilizing the convexity of the dual problem. None of these algorithms, however, provide a closed-form solution to the capacity, because of the complexity of the optimization problem.

In this paper, we establish a closed-form solution for the capacity of the downlink channel for a distributed antenna system. We assume there is a single user that receives the signal from multiple transmit ports, each of which is equipped with an antenna array and a dedicated power budget. The special case of this problem has been studied in \cite{mvu}, where each transmit antenna port has a single antenna. We generalize the results in \cite{mvu} for the case where each antenna port has multiple antennas and the antenna ports are geographically distributed in the cell coverage region.  We consider two cases for availability of the CSI. In the first scenario, we assume that CSI is available both to the transmitters and the receiver and we show that the optimal signaling for each of the antenna ports is beamforming. In the second case, we assume that we have a symmetric fading channel, and the CSI is only available to the receiver. For this case we show that each of the antennas of each port should send independent signals. In other words, we show that the possibility for cooperation among the antenna ports does not help in increasing the capacity, when we have a symmetric fading channel. For both cases, by solving the associated stochastic optimization problem, we establish the capacity in closed form. We also determine the impact upon capacity of interference from other cells, assuming this interference is treated as noise.

Having established this information theoretic analysis of the ergodic capacity of distributed antenna systems, in the second part of this paper we investigate the effect of antenna placement on the performance of such systems. The main reason for looking into the placement problem is the fact that the performance gain of distributed antenna systems (the capacity gain or equivalently the power saving gain) is largely influenced by antenna locations. In \cite{ase}, it was shown that the placement of distributed antennas can significantly influence system performance and that this optimal placement magnifies the advantages of using DAS over the traditional systems with centrally-located antennas~\cite{who}. In other words, the capacity increase of distributed antenna systems is largely influenced by antenna locations. However, many of the studies on the placement optimization of antennas \cite{forcna, forcna3, forcna5} impose restrictions on the topology of the network, such as linear cells or antennas deployed along a circle. To address the problem without imposing these topological constraints, this paper investigates the optimal location of the transmit antennas in distributed antenna systems in a general framework. While our framework can incorporate a broad set of performance metrics, our results will focus on the capacity and power efficiency gains obtained through optimal placement.

The remainder of this paper is organized as follows. First in Section \ref{sec:basic}, we provide a basic description of our distributed antenna system. In Section \ref{sec:cap}, we establish the capacity results for both cases of having channel side information at the transmitter or not. We also analyze the effect of interference from the neighboring cells on our capacity results in Section \ref{sec:interf}. In Section \ref{sec:sol}, after formulating our placement optimization problem, we introduce two different approaches for finding the optimal placement. We investigate some power saving techniques that can be applied to our generalized DAS system in Section \ref{sec:ps}. Section \ref{sec:sim} illustrates the performance of the proposed algorithms in different settings.
Finally, in Section \ref{sec:con} we conclude the paper and present some possible extensions.

\section{System Model}\label{sec:basic}
In this section we describe the system model for analyzing the downlink performance of distributed antenna systems in a cellular setting.
\begin{figure}[t!]	
\begin{center}
\includegraphics[width=\figwww]{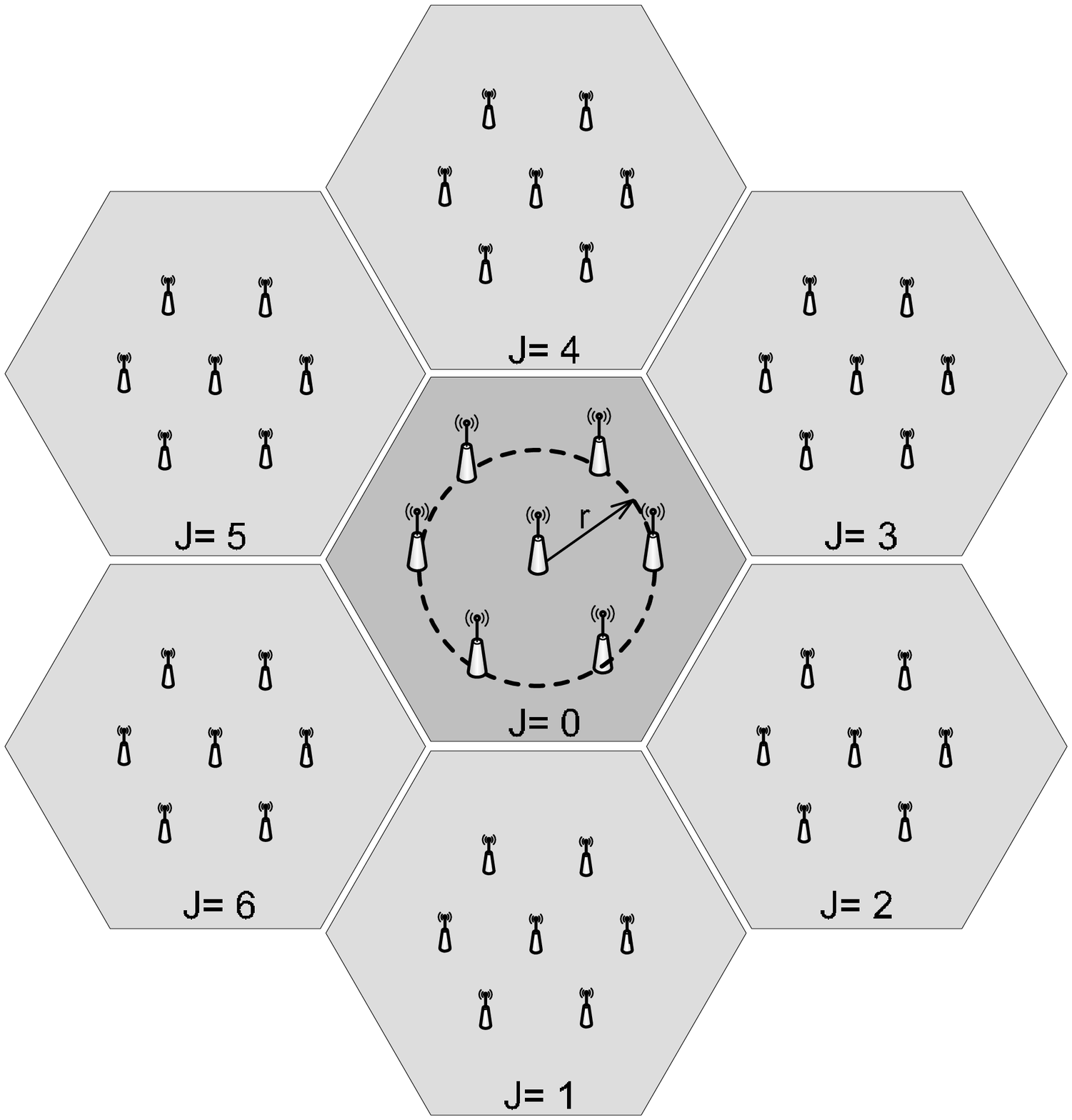}
\end{center}
\caption{Structure of distributed antenna system in a cellular setting}\label{fig:In}
\end{figure}
Figure \ref{fig:In} is the general architecture of the distributed antenna system, $DAS(N,L)$, in the multicell environment we consider in this paper. We assume that each cell is covered with a total number of $N$ distributed antenna ports and each port has L microdiversity antennas. We assume that the antenna ports are all connected to the central antenna port via dedicated links (e.g fiber) with unlimited bandwidth and zero delay. We consider a one-tier cellular structure in which a given cell is surrounded by a continuous tier of six cells that can cause interference. The central cell in Figure \ref{fig:In} is indexed by $j=0$, while the surrounding cells are indexed by $j=1,...,6$. We denote the positions of the ports in the central cell with a $1 \times 2N$ vector $\textbf{P}$ which is defined as
\begin{equation}
	\textbf{P} = [ \textbf{p}_1 , \textbf{p}_2, ... , \textbf{p}_N ], \label{positions}
\end{equation}
where each of the $\textbf{p}_i$'s indicate the position of the $i^\text{th}$ port in the Cartesian plane as $\textbf{p}_i \overset{def}{=} [x_i,y_i]$ for $i \in \{1,2,... ,N \}$. Denoting the centroid of the $j$th cell in the Cartesian plane with $\textbf{o}^j \overset{def}{=} [o_x^j,o_y^j]$, we assume that the relative positions of the ports in the $j$th cell with respect to $\textbf{o}^j$ is identical for all $j=1,2,...,6$. In other words, if we denote the location of the $i^\text{th}$ port in the $j$th cell with $\textbf{p}_i^j$, we assume that for all antenna ports ($i=1,2,...,N$) in the neighboring cells ($\ j=1,2,...,6$) we have\footnote{Note that we put the origin of the Cartesian plane at the centroid of the central cell ($j=0$)} 
\begin{equation}\label{allcells}
\textbf{p}_i^j = \textbf{p}_i + \textbf{o}^j.
\end{equation}

All of the $NL$ antennas in our $N$ distributed antenna ports together construct a macroscopic MISO system with the $NL \times 1$ channel vector
\begin{equation}
	\textbf{h}(j) = [ {h_1{(j)}} , h_2{(j)}, ... , h_N{(j)} ],
\end{equation}
where $ h_n{(j)}$ denotes the channel vector gain from the $L$ antennas of port $n$ to the user at location $\textbf{u}$ which is a function of the distance $r$ between the $n$th transmitter port and the user. Because of the different distances between the antenna ports and the user, our channel vector gain should encompass not only the small scale fading but also the large scale fading together with the path loss as
\begin{equation}\label{channel}
  h_n{(j)} = \sqrt{ \frac{ g^{(j)}_n}{\mathcal{L}^{(j)}_n(r(\textbf{p}_n,\textbf{u}))} } ~  \left[ f^{(j)}_{n,1}, f^{(j)}_{n,2},...,f^{(j)}_{n,L}  \right],
\end{equation}
where $f^{(j)}_{n,i}$ denotes short term fading in the channel at the $i$th antenna of port $n$. In particular, we assume that the fading coefficients are symmetric in the sense that the distribution of $-f^{(j)}_{n,i}$ is identical to the distribution of $f^{(j)}_{n,i}$ for all $1 \leq n \leq N, 1 \leq i \leq L$. One example of such a symmetric fading channel that can be readily applied to our formulation is Rayleigh fading, where $f^{(j)}_{n,i}$ is modeled by a complex Gaussian random variable distributed as $\mathcal{CN}(0,1)$. Also $g^{(j)}_n$ is a random variable representing the shadow fading. $\mathcal{L}^{(j)}_n(r)$ in (\ref{channel}) is the power path loss function which is a function of distance $r$ and also depends on propagation frequency, as we will introduce later. In this paper we consider an exponential path loss function of the form $\mathcal{L}(r) = \beta \times r^{\alpha}$, where $\alpha$ is the path loss exponent whose value is normally in the range of two to six. We assume independent shadow fading between each port and also independent fast fading at each antenna per port. Here, the distance function $r(\cdot)$ can be calculated as
\begin{equation} \label{fas}
r(\textbf{p}_i,\textbf{u}) = \max \{\mathcal{D}(\textbf{p}_i,\textbf{u}) , r_0 \} \ \ \ \  \ i\in \{1,2,... , N \},
\end{equation}
where $\mathcal{D}(\textbf{p}_i,\textit{u})$ is the Cartesian distance between the user at location $\textbf{u} \in \Pi$ and the $i^{th}$ antenna port's location $\textbf{p}_i$ and can be written as
\begin{equation}
\mathcal{D}(\textbf{p}_i,\textit{u}) = \sqrt{ (p_x- u_x)^2 + (p_y-u_y)^2 }.
\end{equation}
 Also, $r_0$ in (\ref{fas}) is the minimum allowable value of $d_i$ for which the far field approximation of the antenna propagation model is valid.
 
Let the transmit signal vector of the $j$th cell be $x{(j)} = \left[ x_1{(j)}, x_2{(j)} ...,x_N{(j)} \right]$, in which $x_n{(j)}$ denotes the $1 \times L$ transmission vector of the $n$th port of the $j$th cell. Then the received signal of the user in the central cell can be written as
\begin{equation}
y = \textbf{h}(0)^T x{(0)} + \sum_{j=1}^6 \sqrt{\gamma_j} ~ \textbf{h}(j)^T x{(j)} + n, = \textbf{h}(0)^T x{(0)} + z,\label{signal2}
\end{equation}
where $n$ is an additive Gaussian noise with variance $\sigma^2_n$ and $\gamma_j$ is a positive real multiuser coding gain that quantifies the fact that the signals from the j$th$ cell are not completely orthogonal to the signals of the central cell\footnote{This coding gain can be set to one if we do not use orthogonal codes in the neighboring cells. On the other hand, $\gamma_j = 0$ indicates that we have no interference from the neighboring cells. Also in a CDMA system, this can quantify the spreading code gain. }. In deriving (\ref{signal2}), since the neighboring cells are sending independent messages, we approximated the sum of interfering signals plus noise as a complex Gaussian random variable $z$ with variance $\sigma^2_z$. We also assume that each port has a separate transmit power budget of $S_n ~ (n=1,2,...,N)$, i.e
\begin{equation}\label{powcon}
\mathbb{E}\left[ x_n{(j)} x_n{(j)}^H \right] = S_n,
\end{equation}
for all ports $1\leq n \leq N$, and in each of the cells $0\leq j \leq 6$. Note that in (\ref{powcon}), the $n$th transmission port can allocate its own power budget $S_n$ among it's antennas, but different ports have no ability for allocating power amongst themselves. We also denote the transmission power constraint of all ports by the vector $\bar{S} \overset{def}{=} \left[ S_1,S_2,...,S_N \right]$.

\section{Downlink Capacity of Distributed Antenna System}\label{sec:cap}
\subsection{On downlink ergodic capacity}
Given the channel model in Section \ref{sec:basic}, for fixed port location matrix $\textbf{P}$ and fixed user's location $\textbf{u}$, we can write the average ergodic rate of a single user under different circumstances. In this section, we consider two cases of channel information at the transmitter: the channel coefficient vector $\textbf{h}(j)$ is  known to the transmitters of the $j$th port (channel with CSIT) or not (channel with no CSIT). In both cases, we assume that the channel coefficient vector is known at the receiver, which is a good approximation for practical systems with receiver channel estimation. The capacity of this channel depends on the power constraint on the input signal vector $x$ in (\ref{signal2}) and also on the availability of the CSI at the transmitter. The goal of this section is to investigate the downlink average ergodic capacity of a user in the following scenarios for different sets of information that are available at the transmission ports.

\subsubsection{CSI only at the receiver}\label{sec:nocsit}
In this case, we assume that the channel vector $\textbf{h}$ is unknown to the transmitter. Because of the Gaussian noise and known channel at the receiver, the optimal input signal is Gaussian with zero mean \cite{dow,mvu}, and the achievable average transmission rate of the user for this case can be found by solving the following stochastic optimization problem:
\begin{align}\label{rateg}
\text{Max : ~~} & \mathbb{E}_{f} \left[ \log_2 \left( 1 + \frac{1}{\sigma_z^2} ~ \textbf{h}(0) ~ Q ~ \textbf{h}(0)^H ) \right) \right] \\
\text{Subject to : ~~} &Q \succeq 0 ~,  \textbf{tr}\left(q_{nn}\right) \leq S_n, ~~~ \forall 1\leq n \leq N ~, \notag 
\end{align}
where $\mathbb{E}_{f} \left[ \cdot \right]$ denotes the expectation with respect to both slow and fast fading, $Q = \mathbb{E}\left[ x{(0)}^H x{(0)} \right]$ is the $NL \times NL$ Hermitian covariance matrix of the Gaussian input of the central cell, and $q_{mn}$ is the part of the covariance matrix $Q$ that corresponds to the correlation of the transmitted signal of the $m$th and the $n$th ports, i.e. $q_{mn} = \left[ x_m{(0)}^T  x_n{(0)} \right]_{L\times L}$. Note that the constraint $\textbf{tr}\left(q_{nn}\right) \leq S_n$ corresponds to the power constraint of each antenna port defined in (\ref{powcon}).  The main issue in solving (\ref{rateg}) is finding the optimal covariance matrix $Q$. Since the per-port power constraint $\textbf{tr}\left(q_{nn}\right) \leq S_n$ is not equivalent to a constraints on eigenvalues of $Q$, the analysis of \cite{dow} cannot be applied here. In other words, after the eigenvalue  decomposition of $Q = U~\Lambda~U^T$, although $\textbf{h}(0)~U$ has the same distribution as $\textbf{h}(0)$ , the partial constraints on the eigenvalues of $\Lambda$ would not be the same as the constraint on (\ref{rateg}). Therefore, the problem is no longer
equivalent to the analysis of the fading channels in \cite{dow} through eigenvalue decomposition. However, inspired by the work in \cite{mvu} and based on the symmetric distribution of the fading channel, we can show that the optimal solution of (\ref{rateg}) is $Q_{csir}^* = diag(\frac{S_1}{L} \textbf{I}_{L},\frac{S_2}{L} \textbf{I}_{L},...,\frac{S_N}{L} \textbf{I}_{L})$, where $\textbf{I}_L$ denotes an $L \times L$ identity matrix. This optimal solution means that each antenna port should transmit independent signals at full power.  Somewhat surprisingly, our optimal transmission strategy in this case is similar to the transmission strategy of a multiple access channel. In other words, for a symmetric fading channel, e.g. a Rayleigh fading channel, the possibility for cooperation among the antenna ports does not help in increasing the capacity. In this case the capacity is
\begin{equation}
C(\textbf{u}, \ \textbf{P}, \bar{S}) = \mathbb{E}_{f} \left[ \log_2 \left( 1 + \frac{1}{\sigma_z^2}  \textbf{h}(0)  Q_{csir}^*  \textbf{h}(0)^H ) \right) \right]  = \mathbb{E}_{f} \left[ \log_2 \left( 1 + \frac{1}{L  \sigma_z^2}  \sum_{n=1}^N \| \textbf{h}_n(0) \|^2  S_n  \right) \right],\label{ic}
\end{equation}
where $\|.\|$ denotes the ${l}^2$-Norm.

\subsubsection{CSI both at the transmitter and the receiver}\label{sec:csit}
In this part, we assume that the channel vector gain $\textbf{h}$ is known at the transmitter. Similar to the case of having CSIR, the optimal input signal is still Gaussian with zero mean, because of the Gaussian noise and known channel at the receiver. The capacity then can be found by solving the following optimization problem
\begin{align}\label{rateg2}
\text{Max : ~~} &  \log_2 \left( 1 + \frac{1}{\sigma_z^2} ~ \textbf{h}(0) ~ Q ~ \textbf{h}(0)^H ) \right)  \\
\text{Subject to : ~~}   &Q \succeq 0 ~,  \textbf{tr}\left(q_{nn}\right) \leq S_n, ~~~ \forall 1\leq n \leq N ~. \notag 
\end{align}
In \cite{mvu}, the problem is solved for the special case of $L=1$ by relaxing the positive semi-definite constraint $Q \succeq 0$. Here, we find the solution for a general $DAS(N,L)$ system. We show in Appendix \ref{app:capcsit} that the optimal covariance matrix $Q_{csit}^*$ is
\begin{equation}
Q_{csit}^* = \left[ q_1, q_2, ..., q_N \right]^H \left[ q_1, q_2, ..., q_N \right],
\end{equation}
where $q_n \overset{def}{=} \frac{\textbf{h}_n^*(0)}{\| \textbf{h}_n(0) \|} \sqrt{S_n}, ~~ 1 \leq n \leq N$. Also, the optimal signaling solution for the $n$th antenna port is beamforming with a beam weight of $\textbf{h}_n^*(0) / \| \textbf{h}_n(0) \|$. As we can see, in each antenna port, the power allocation among $L$ antennas is proportional to the amplitude of the channels, while the phases are matched and the total transmit power is set to its maximum allowed budget. For the special case of $L=1$, each antenna port should match the phase of its associated channel so that the signals add coherently at the receiver and also set the amplitude of the transmitted signal independent of the channel and fixed according to the power constraint. In other words, the key information that we need for achieving capacity is the phase of the channel gains, but not the amplitudes. Note that  for the case of $L=1$, there is actually no power allocation among the transmit antennas and the transmit power of the $n$th antenna port is fixed at $S_n$. On the other hand, if we only have a single antenna port ($N=1$) with $L$ microdiversity antennas, beamforming with the weights that are proportional to the channel gains is essential to achieve the maximum capacity.

The average ergodic rate of DAS(N,L) system with CSIT can then be written as
\begin{align}
C(\textbf{u},& \ \textbf{P}, \bar{S}) = \mathbb{E}_{f} \left[ \log_2 \left( 1 + \frac{1}{\sigma_z^2} ~ \textbf{h}(0) ~ Q_{csit}^* ~ \textbf{h}(0)^H ) \right) \right]  \notag \\
&= \mathbb{E}_{f} \left[ \log_2 \left( 1 + \frac{1}{\sigma_z^2} \left( \sum_{n=1}^N \| \textbf{h}_n(0) \| \sqrt{S_n} \right)^2 \right) \right],\label{bf}
\end{align}
where $\mathbb{E}_{f} \left[ \cdot \right]$ denotes the expectation with respect to both slow and fast fading. Comparing (\ref{bf}) and (\ref{ic}), we see that under the per-port power constraint, the presence or lack of channel phase information at the transmitter has a significant impact on the optimal transmit strategy and hence on the channel capacity.

\subsection{Interference from neighboring cells}\label{sec:interf}
We also investigate the effect of interference from the neighboring cells on ergodic capacity of the system both with and without CSI. We assume there is no cooperation between cells and that intercell interference is treated as noise. As a result, regardless of the availability of CSIT in a given cell, the interfering signals from neighboring cells add incoherently at each user's receiver. Therefore, we can write the power of interference plus noise (complex Gaussian random variable $z$ in (\ref{signal2})) for a user in the central cell ($j=0$), as
\begin{equation}\label{int_noise}
\sigma_z^2 = { \sum_{j=1}^6  \sum_{i=1}^N  \frac{{ \gamma_j } }{\mathcal{L}(r(\textbf{p}^{(j)}_i,{\textbf{u}}))} {S}  + \sigma^2_n}, 
\end{equation}
where we assume that all antenna ports in all neighboring cells are using the same transmit power of ${S}$. Therefore, by replacing $\sigma_z^2$ with the variance of the noise $\sigma_n^2$ in (\ref{ic}) or (\ref{bf}), we can consider the effect of interference on the expected ergodic capacity. 

\begin{figure}[t]		
\begin{center}
\includegraphics[width=\figwww]{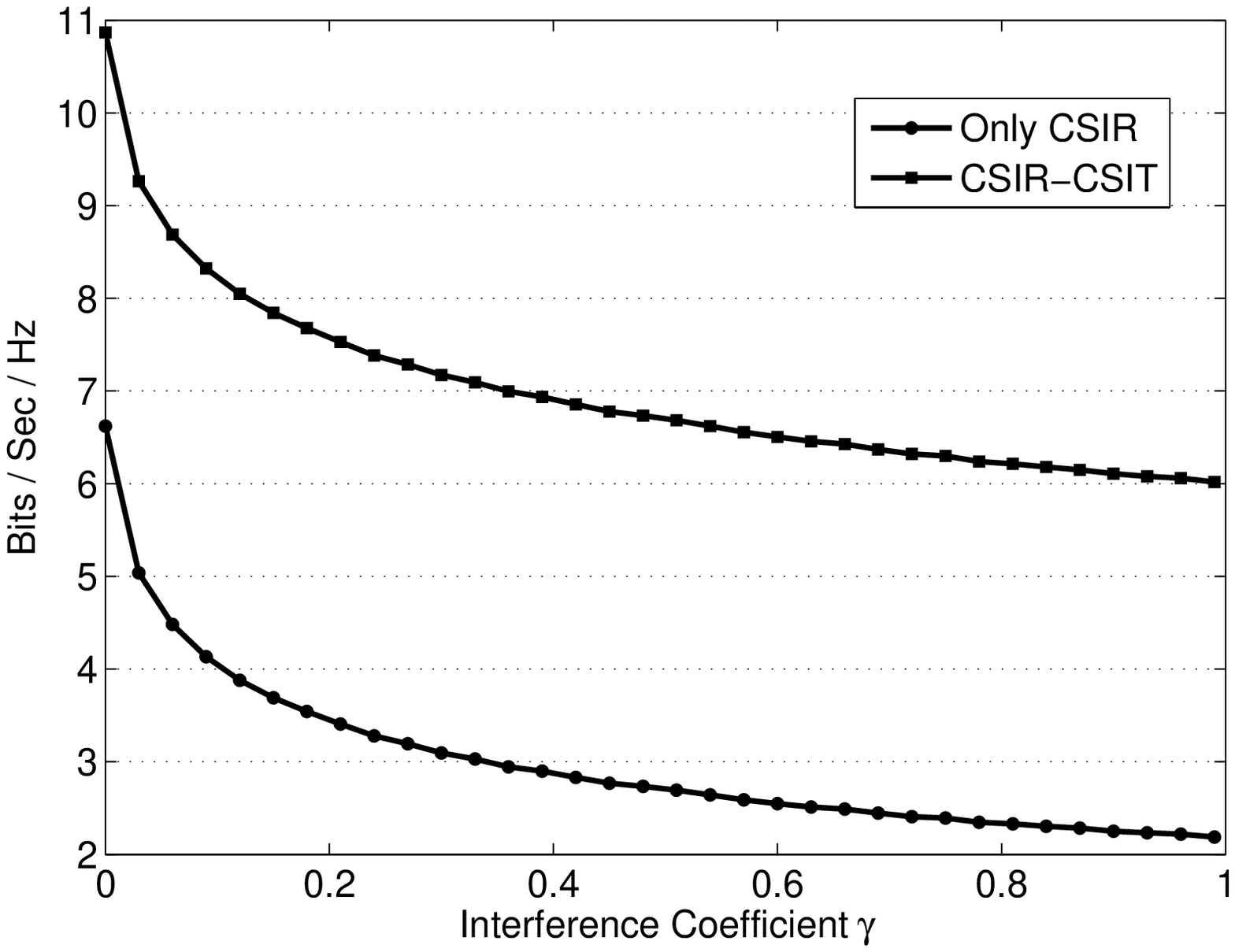}
\end{center}
\caption{Average ergodic rate of a user, randomly placed in the coverage region, as a function of }\label{fig:lm_}
\end{figure}

In order to illustrates the effect of interference on the downlink capacity of $DAS(7,1)$, we consider the following model. As shown in Figure \ref{fig:In}, we assume that one port is in the center and six other ports are in a circle of radius $r = R/2$ around it, where $R$ is the radius of the hexagonal coverage region.  We assume independent log-normal shadowing of variance $\sigma_{sh} = 8$ between each port and the user. The path loss exponent is assumed to be $\alpha = 4$ and we set the power constraint on each port\footnote{in absence of any interference from other ports} such that the SNR of a user at distance $R$ is $10$ dB. Figure \ref{fig:lm_} illustrates the downlink average ergodic rate of a user, averaged over both the shadowing distribution and user location which is assumed to be uniformly distributed inside the cell. We consider both cases where CSI is only available to the receiver and where CSI is available both to the transmitter and the receiver. As expected, having the CSI at the transmitter improves the average ergodic rate substantially. It is also worth mentioning that in the case of having only CSIR, the SINR would be much worse for the users on the edge of the cell in comparison with the case where we have CSIT (since the interference terms will add incoherently even when we have CSIT, whereas the signal is adding coherently).  

\section{Antenna Placement}\label{sec:sol}
In this section we focus on the placement optimization of the antenna ports in $DAS(N,L)$. As we will show later in Section \ref{sec:sim}, depending on different assumptions about the availability of CSI, the optimal antenna placement can have a huge impact on the energy efficiency of distributed antenna systems. For the sake of simplicity, we neglect the effect of fast fading in our optimization problem, i.e we replace the fast fading terms in the channel gain vector (\ref{channel}) with $f_{n,i} = 1,$ for all $1\leq n \leq N ,~ 1 \leq i \leq L$. In addition, we start with the scenario when there is no interference and the metric that we want to optimize is the cell averaged ergodic capacity.  The averaging here is done with respect to the location of the user, which is uniformly distributed within the cell region. We use $\textbf{u}$ to denote the position of the user. The  ergodic capacity in $bit/s/Hz$ conditioned on the user's position $u \in \Pi$, antenna placement $\textbf{P} \in \Pi$ and also the power budget $\bar{S}$ is introduced in Section \ref{sec:cap} for two different cases where the CSIT is available at the transmitter or not. Specifically, depending on the use of (\ref{ic}) or (\ref{bf}) for the capacity, the cell averaged ergodic capacity $\bar{C}$ for a particular placement of the antennas can be calculated by averaging $C(\textbf{u},\textbf{P},\bar{S})$ over user positions $u$ inside the cell region $\Pi$ as
\begin{equation}\label{ratee}
\bar{C}(\textbf{P}, \bar{S}) = \mathbb{E}_{u}  \ \  C(u, \ \textbf{P}, \bar{S}) ,
\end{equation}
where $\mathbb{E}_{u} \{ \cdot \}$ denotes the expectation with respect to the location of the user and in our model we assume that it has a uniform distribution inside the cell region.

Now we can describe the placement optimization problem for DAS that we consider in this section. 
\begin{problem}\label{problem1}  Consider N to be a positive integer quantifying the number of ports in our distributed antenna system. Let the variance of the log-normal shadowing $\sigma_{sh}$ and the minimum distance for far-field approximation $r_0$ be given design parameters. Also assume that the power path loss function $\mathcal{L}(\cdot)$ is known and we have the same power budget at each antenna port, denoted by ${S}$, i.e $\bar{S} = \left[ S,S,...,S \right]$. We want to find the solution to the following optimization paradigm:
\begin{equation}
\label{pr1}
\textbf{P}^* = \text{arg}\min_{\textbf{P}}  S,
\end{equation}
subject to :
\begin{equation}
\label{ct1}
\bar{C}(\textbf{P}, \bar{S})  \geq C_{t},
\end{equation}
\begin{equation}
\label{ct2}
\textbf{p}_i  \in \Pi \  \ \ \ \  \ i\in \{1,2,... , N \},
\end{equation}
where $C_{t}$ is the target for the expected cell average ergodic rate. Here $\text{P}$, as defined in Section \ref{sec:basic}, is the matrix containing the collection of locations for all the ports in the system.
\end{problem}

\begin{remark} \label{rem1}
It can be easily seen that Problem \ref{problem1} is equivalent to maximizing the expected cell average ergodic rate with a constraint on the maximum power budget ${S}$.
\end{remark}

In the remainder of this section, we propose two different solutions for the aforementioned placement optimization problem. First we present a simple approximation method that casts the problem in a simpler space that is independent of channel characteristics. Next we propose a very general framework for placement optimization that can handle more complex optimization metrics.


\subsection{Simple approximation method}\label{sec:na}
Either of the ergodic rate formulas (\ref{ic}) or (\ref{bf}) introduced in Section \ref{sec:cap} involve a sum over $NL$ nonnegative variables among which the $L$ terms of the antenna port with the smallest distance $r$ dominates the sum. One approach for tackling the placement optimization problem is to only consider the dominant terms in the summation and ignore the rest. Starting with (\ref{ic}) where the coherent transmission is not possible due to the lack of phase information at the transmitter, by considering the closest antenna port to the user we can approximate the average capacity with the following lower bound \cite{forcna}
\begin{align}\label{lb}
\bar{C}(\textbf{P},\bar{S}) \geq \mathbb{E}_{u} \ \  \mathbb{E}_{f} \log_2 \left( 1 +   \frac{ g_m ~ S}{\mathcal{L}(r_{min})}  \right),
\end{align}
where
\begin{align}
r_{min} &= \min_{1 \leq i \leq N} \  \mathcal{D}(\textbf{p}_i,\textbf{u}), \label{rmin}  \\
m &= \arg \min_{1 \leq i \leq N}  \mathcal{D}(\textbf{p}_i,\textbf{u}).
\end{align}
When CSIT is available, the approximations in (\ref{lb}) remain unchanged, except the fact that the power term $S$ should be multiplied by $L$, because of the coherent summation of the signals at the receiver. However, this constant $L$ has no effect on our placement optimization problem so we continue with (\ref{lb}) as a lower bound on average capacity. Also, note that the approximation in (\ref{lb}) works better for higher values of path loss exponent $\alpha$, since the larger values of $\alpha$ magnify the difference between the dominant term and the second largest value in the summations of (\ref{ic}) and (\ref{bf}). By exchanging the order of the two expectation operators and noting that the function $f(x) = \log_2(1 + \frac{a}{x}) $ $(a > 0)$ is a convex function, we can apply Jensen's inequality to obtain a lower bound for averaged capacity as follows
\begin{equation}
\bar{C}(\textbf{P},\bar{S})  \geq \mathbb{E}_{f} \ \ \mathbb{E}_{u} \log_2 \left( 1 +  \frac{ g_m ~ S}{\mathcal{L}(r_{min})}   \right)  \geq \mathbb{E}_{f} \ \  \log_2 \left( 1 +    \frac{ g_m ~ S}{  \mathbb{E}_{u} \{ {\mathcal{L}(r_{min})} \}  }  \right). \label{shad}
\end{equation}
We can use a similar approximation, based on Jensen's inequality, to average over shadowing. To this end, we define the auxiliary random variable $\tilde{g}_m \overset{def}{=} \log_{10} g_m$ and rewrite (\ref{shad}) as
\begin{align}
\bar{C}(\textbf{P},\bar{S}) & \geq   \mathbb{E}_{f} \ \  \log_2 \left( 1 +    \frac{{S}  \  10^{\tilde{g}_m}}{  \mathbb{E}_{u} \{ {\mathcal{L}(r_{min})} \}  }  \right) \label{shad3} \\
& \geq   \ \  \log_2 \left( 1 +    \frac{{S}   \  10^{ \mathbb{E}_{f} \{ \tilde{g}_m \} }  }{  \mathbb{E}_{u} \{ {\mathcal{L}(r_{min})} \}  }  \right) \label{shad4} \\
& =   \ \  \log_2 \left( 1 +    \frac{{S} \  }{  \mathbb{E}_{u} \{ {\mathcal{L}(r_{min})} \}  }  \right). \label{shad5}
\end{align}
Recall that, according to our model, the shadowing term $g_m$ has a log-normal distribution. Hence, $\tilde{g}_m$ in (\ref{shad4}) has a zero mean normal distribution. Also, the inequality in (\ref{shad4}) is a direct application of Jensen'e inequality to (\ref{shad3}) by noting that the function $\log ( 1 + a \ 10^x )$ is a convex function of $x \in \mathbb{R}$, for any positive coefficient $a$. Note that equations (\ref{shad3}), (\ref{shad4}) and (\ref{shad5}) suggest that we can ignore the effect of shadowing and still obtain an upper bound on the minimum power required to achieve a certain average ergodic rate of the cell. We will confirm this observation in Section \ref{sec:sim} via simulation, in which we illustrate that shadowing has a negligible effect on the placement optimization of the ports.

Now, in order to obtain an upper bound for the solution of Problem \ref{problem1}, we can replace constraint (\ref{ct1}) with a more conservative constraint of the type
\begin{equation}
  \log_2 \left( 1 +    \frac{{S}   }{  \mathbb{E}_{u} \{ {\mathcal{L}(r_{min})} \}  }  \right) \geq C_{t} \ . \label{csj}
\end{equation}
This inequality indicates that the approximate solution\footnote{note that replacing (\ref{ct1}) with (\ref{csj}) will shrink the feasible set of Problem \ref{problem1} and hence give an upper bound on the solution } of Problem \ref{problem1} can be obtained by minimizing $ \mathbb{E}_{u} \{ {\mathcal{L}(r_{min})} \}$. Therefore, with this lower bounding we can get a criterion for antenna location design which is to minimize the expectation of the path loss function $\mathcal{L}$ (for any such function) from the randomly distributed user to the nearest antenna port.
The following example illustrates this bounding approach for a simple example, where we want to optimally place three ports within a circular coverage region.


\begin{example}\label{ex:1}
Let $r_0$ be zero, the number of ports $N = 3$ and the cell region be a hexagon of radius $R$, as illustrated in Figure \ref{fig:e3}. The goal is to find the approximate solution of Problem \ref{problem1} for this setting. Due to the symmetry of the problem, one can argue that the optimal placement of these three single-antenna ports should also be symmetric around the center of the region. One possibility for such a symmetric placement is illustrated in Figure \ref{fig:e3} where the regions separated by dashed lines are sectors with central angles of $2 \pi / 3$ and the antenna ports are located on the bisector of the central angle. Here, for a randomly placed user, $r_{min}$ in (\ref{rmin}) would be the distance between this user and the associated port in its sector. Now, in order to find the optimal location of the antenna ports, we need to minimize $ \mathbb{E}_{u} \{ {\mathcal{L}(r_{min})} \}$ or equivalently $ \mathbb{E}_{u} \{ r_{min}^\alpha \}$, where the expectation is over a randomly placed user $u$ within the sector.  The solution is the centroid of the sector. The centroid here depends on the path loss exponent $\alpha$ (for instance, for $\alpha = 2$ this reduces to the geometric centroid). Using this lower bound approximation, the optimal radius $r^*$ to deploy the antenna ports ranges from $r^* = 0.57~R$ to $r^* = 0.58~R$, depending on the path loss exponent $\alpha$ (see Figure \ref{fig:ns}). This indicates that, using this lower bound approximation approach, the optimal position does not change much for different values of $\alpha$. Note that this solution is based on the aforementioned approximation approach and has some implicit assumptions and approximations that make it suboptimal.
\begin{figure}
 \begin{center}
  \includegraphics[width=\figwwww]{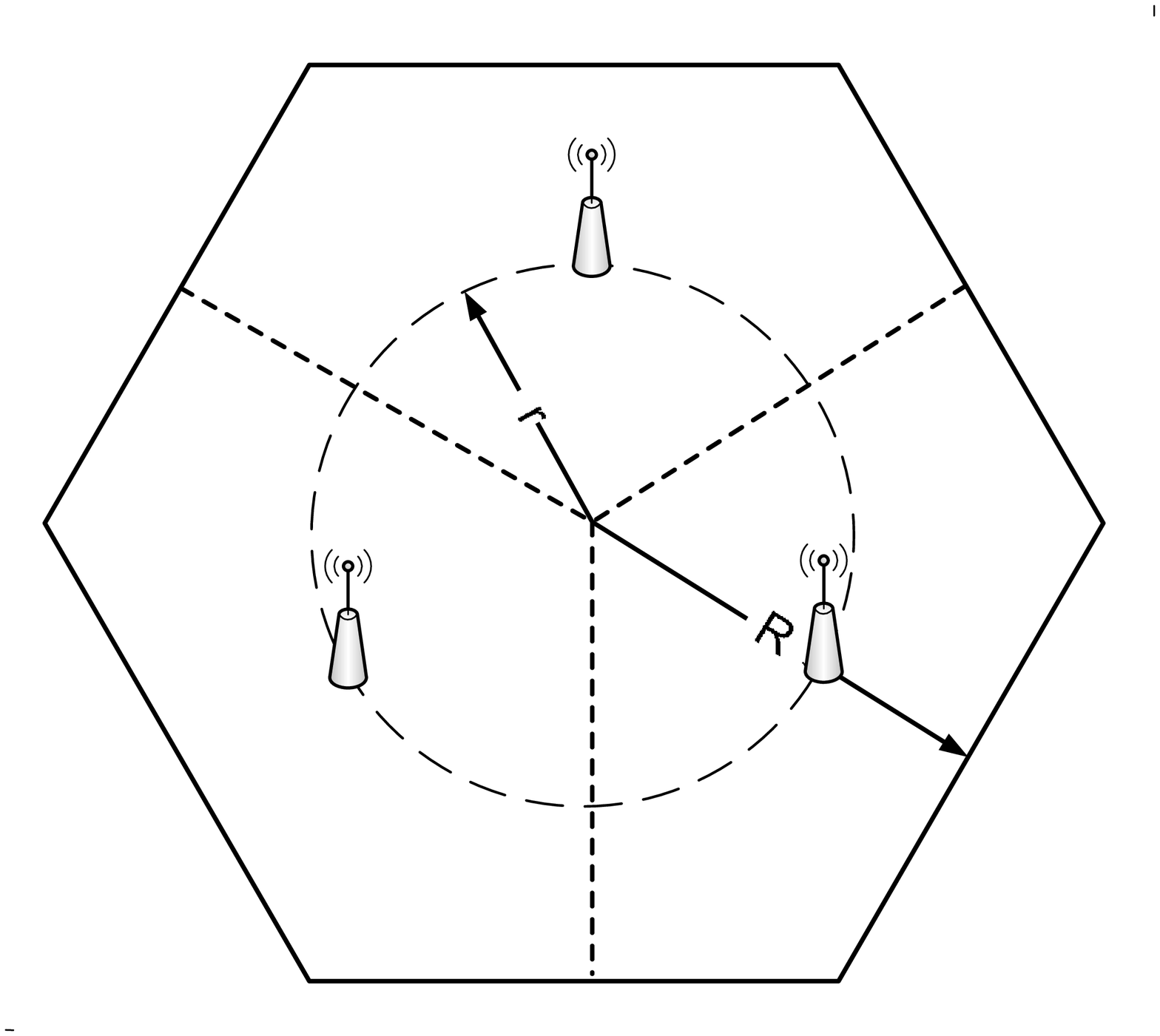}
  \end{center}
  \caption{Three port placement within a hexagonal region of radius R. Location of the transmitter ports are denoted by star shapes.}\label{fig:e3}
\end{figure}

\end{example}

As we can see, the antenna port locations obtained with this approach only depend on the antenna port number $N$, the user distribution (including cell shape), and the path loss exponent $\alpha$. However, these optimal locations are completely independent of the availability of the CSI at the transmitter. In other words, using either (\ref{ic}) or (\ref{bf}) for the ergodic rate yields the same solution. Moreover, the solution is also not affected by the underlying shadowing model in the system. Note that the proposed criterion aims to maximize a lower bound of the performance metric $\bar{C}$, rather than $\bar{C}$ itself. Therefore, the obtained antenna port locations are suboptimal. It is also not possible to generalize this model in order to take into account the interference from other users within the cell or from neighboring clusters. The other problem is that this approach has an implicit assumption that in the optimal placement, all the ports would be placed far away from each other so that the lower bound (\ref{lb}) is tight, which might not be true in general. However, this method becomes more accurate for the regime where we have high path loss since the lower bound (\ref{lb}) becomes tighter.


In the next subsection, we propose a more general technique based upon stochastic approximation for optimizing the placement, that can handle a much more general class of objective functions.

\subsection{Stochastic Update of the Locations}\label{sec:sap}
In this part, based on stochastic approximation theory, we introduce a general framework for solving the placement optimization paradigm. For the sake of clarity, rather than solving (\ref{pr1}) directly, we will first focus on the equivalent problem of maximizing the average ergodic rate with the given power budget $\bar{S}$. We touch upon the generalization of the proposed algorithm for more sophisticated optimization metrics and also the direct solution to (\ref{pr1}) afterwards.

Assume we want to find the optimal placement vector $\textbf{P}$ in order to maximize
\begin{equation}
\bar{C}(\textbf{P},\bar{S}) =  \mathbb{E}_{u} \ \ \mathbb{E}_{f}  \ \ \ C(\textbf{u},  \textbf{P}, \textbf{g} ,\bar{S}), \label{ec}
\end{equation}
where $\textbf{g} \overset{def}{=} [g_1, g_2, ... , g_N ] $ is defined to be a random vector of size $N$, indicating the shadowing and $\mathbb{E}_{f} [\cdot]$ denotes the expectation with respect to this shadowing. Also $C({\textbf{u}}, \textbf{P}, \textbf{g},\bar{S})$ is the instantaneous ergodic rate for a given realization of the user position ${\textbf{u}}$, shadowing vector ${\textbf{g}}$, and given location matrix ${\textbf{P}}$, depending on which information we assume to be available at the transmitter. In particular, when we assume that CSI is only available to the receiver but not the transmitter, from (\ref{ic}) we can write the capacity as
\begin{equation}\label{capas1}
C({\textbf{u}}, {\textbf{P}},{\textbf{g}},\bar{S}) \overset{def}{=} \log_2 \left( 1 + \frac{{S}}{\sigma^2_n}  \sum_{n=1}^N  \frac{{g_n} }{\mathcal{L}_n(r(\textbf{p}_n,{\textbf{u}}))}   \right),
\end{equation}
and when CSI is assumed to be known both at the transmitter and the receiver as in (\ref{bf}), the instantaneous capacity can be written as
\begin{equation}\label{capas2}
C({\textbf{u}}, \textbf{P}, {\textbf{g}},\bar{S}) \overset{def}{=} \log_2 \left( 1 + L \frac{{S}}{\sigma^2_n} \left( \sum_{n=1}^N \sqrt{ \frac{{g_n} }{\mathcal{L}_n(r(\textbf{p}_n,{\textbf{u}}))}  }  \right)^2 \right).
\end{equation}

 Now, in order to find the optimal location matrix $\textbf{P}^*$ that maximizes (\ref{ec}), we use the Robbins-Monro procedure \cite{135} from stochastic approximation theory. Before getting into the algorithm that finds $\textbf{P}^*$, we will briefly describe the Robbins-Monro procedure. In stochastic approximation, maximizing a utility function $U(\textbf{x})$ can be done via an iterative update of the optimization variable $\textbf{x}$ in the following way
\begin{equation}\label{sap1}
    \textbf{x}_{t+1} = \textbf{x}_t + \sigma^t \left( U_x(\textbf{x}_t) + M_t \right),
\end{equation}
where $U_x(\textbf{x}) \overset{def}{=} \frac{ \partial U}{\partial \textbf{x}}$, $M_t$ is some zero mean noise in the estimate of $U_x(\textbf{x})$ and $\{ \sigma^t \}_{t = 1}^\infty $ is a square summable but not summable sequence.\footnote{The convergence of this Robbins-Monro iteration is shown under broad conditions in and \cite[chapter 6]{135}. In particular, the iterations converge to a local optimum when the noise terms $M_t$ have bounded variance.} In our specific placement optimization problem, we can look at the expected ergodic capacity as the utility function that we want to maximize and use the Robins-Monro procedure to update the location of the ports at each iteration.

We can now specify the following iterative algorithm for solving (\ref{ec}).
\begin{enumerate}
\item Initialize the location of the ports randomly inside the coverage region and set $t = 1$
\item Generate one realization of the shadowing vector $\textbf{g}_t$,\footnote{We can also consider the fast fading vector $\textbf{f}_t$ in our gradient estimation} based on the probabilistic model that we have for the shadowing of the channel.
\item Generate a random location $\textbf{u}_t$ for the position of the user based on the distribution of the users in the coverage area.
\item Update the location vector as
\begin{equation}\label{s4}
\textbf{P}_{t+1} = \textbf{P}_t +  \sigma^t \ \frac{ \partial C(\textbf{u}_t,  \textbf{P}, \textbf{g}_t,\bar{S}) }{\partial \textbf{P}} \big|_{ \textbf{P}_t }
\end{equation}
\item let $t = t + 1$ and go to step 2 until convergence.
\end{enumerate}

In order to provide some intuition about this algorithm, we will try to put it in the form of the Robins-Monro recursion (\ref{sap1}). As we can see, $\frac{ \partial C(\textbf{u}_t,  \textbf{P}, \textbf{g}_t ,\bar{S})) }{\partial \textbf{P}} $ is an unbiased estimator of $\mathbb{E}_{u} \  \mathbb{E}_{f} \  \frac{ \partial C(u, \ \textbf{P},\textbf{g}) }{\partial \textbf{P}} $. Hence by defining
\begin{align}
M_t &= \frac{ \partial C(\textbf{u}_t,  \textbf{P},\textbf{g}_t,\bar{S}) }{\partial \textbf{P}} \label{bistt} - \mathbb{E}_{u}   \mathbb{E}_{f} \left[  \frac{ \partial C(\textbf{u},  \textbf{P}, \textbf{g},\bar{S}) }{\partial \textbf{P}} \right] \\
	&=    \frac{ \partial C(\textbf{u}_t,  \textbf{P},\textbf{g}_t,\bar{S}) }{\partial \textbf{P}} -  \frac{ \partial  }{\partial \textbf{P}} \{ \mathbb{E}_{u}   \mathbb{E}_{f} \left[ C(\textbf{u},  \textbf{P},\textbf{g},\bar{S}) \right] \},  \label{bist}
\end{align}
we can rewrite the position update (\ref{s4}) of step 4 in the form of the Robins-Monro recursion by noticing that in our placement problem the utility function $U(\cdot) = \mathbb{E}_{u}  \mathbb{E}_{f}C(\cdot)$ and the zero mean noise term $M_t$ is defined in (\ref{bist}). 

Now we describe how we can solve (\ref{pr1}) directly, rather than solving the equivalent problem of maximizing the average ergodic rate with given power budget ${S}$. Similar to \cite{dan2}, a primal-dual algorithm can be used to solve (\ref{pr1}) with the same stochastic update mechanism. Briefly, the dual problem and the Lagrange multiplier can be updated in an iterative fashion with a stochastic gradient method \cite{dan2}. Unfortunately, due to the non-convexity of the objective function, there is no guarantee that the algorithm converges to the globally optimum solution, and it's possible that stochastic gradient steps may converge to a local optimum. This problem has been addressed for the special case where the path loss function has an exponential fall off with distance, where we showed the placement optimization problem can be reformulated as a convex optimization problem \cite{ms}.

In general, we do not have to restrict ourselves to the expected ergodic capacity metric $ \mathbb{E}_{u}  \mathbb{E}_{f}C(\cdot)$ in (\ref{ec}) and it can be replaced with other metrics such as outage probability. This can be done simply by replacing $C(\cdot)$ in $(\ref{s4})$ with any arbitrary performance metric that we want to optimize. In particular, in the following part, we show how the interference from the neighboring cells can be considered in the antenna port placement optimization.

\subsection{Interference Effect}\label{sec:int}
In this part, we investigate the effect of interference from the neighboring cells on antenna port placement optimization. As we mentioned in Section \ref{sec:interf}, by replacing $\sigma_z^2$ from (\ref{int_noise}) with the variance of the noise $\sigma_n^2$ in (\ref{capas1}) or (\ref{capas2}), we can consider the effect of interference on the expected ergodic capacity. The same stochastic update algorithm of Section \ref{sec:sap} can be applied here for placement optimization of antenna ports, simply by replacing the capacity formula in $(\ref{s4})$. The only difference here is the assumption that we have about the location of the ports in the neighboring cells. As we described in Section \ref{sec:basic}, we assume a similar relative position of all antenna ports in each cell with respect to the center of that cell. Therefore, by updating the location vector in (\ref{s4}), the location of all interfering ports would also change accordingly. This indeed forces the final optimal layout of the antennas (after convergence of the algorithm) to be identical in all the cells $j=0,1,...,6$. As we will show in the next section, the effect of interference in the optimal location of the antenna ports is to move the ports towards the center of the cell in order to minimize the power of the antenna ports outside the cell boundaries.

\section{Power saving techniques}\label{sec:ps}
In this section we study some power saving techniques that can be applied to our generalized DAS system. In particular, in the first part, we study the optimal output power of the antenna ports with respect to their position in the cell in order to mitigate the effect of interference. In the second part, we analyze the single transmission strategy where only one antenna port is used in the downlink. We also generalize the stochastic gradient method of Section \ref{sec:sap} in order to find the joint optimal location and power allocation of the antenna ports for the single transmission strategy.

\subsection{Distribution of power}
Until now, we always assumed that we have the same power constraint ${S}$ on the output power of antenna ports in the cell. The alternative is to adjust the power constraints of the ports according to their position in the cell. In other words, instead of having a fixed power vector $\bar{S} = \left[ {S}, {S},...,{S} \right] $, we can treat the power budget of each antenna port $S_n$ $(1 \leq n \leq N)$ as an optimization variable. In order to have a fair comparison with the case that all of the output powers are fixed to be $\bar{S}$, we need to put a constraint on the total consumed power in the cell as $\sum_{i=1}^N s_i = N \times {S}$. Hence the averaged ergodic rate maximization problem becomes
\begin{align}\label{pd}
 \text{Maximize} &: \bar{C}(\ \textbf{P}, \bar{S}) \\
 \text{Subject to} &: \sum_{n=1}^N S_n \leq N \times {S}, \notag
\end{align}
where $  \textbf{P}$ and $ \bar{S}$ are optimization variables. A simple variation of the stochastic gradient method of Section \ref{sec:sap} can handle this constrained optimization problem ( The algorithm is not included due to space limitations.) 
As we will observe in the numerical results of the next section, by increasing the effect of interference (increasing $\gamma$) the optimal power allocation algorithm tends to dedicate more power to the central antenna ports in order to reduce the amount of leaking power to the neighboring cells that cause interference.

\subsection{Single transmission strategy}\label{sts}
Several transmission strategies are possible in a distributed antenna system \cite{selection}. Until now, we assumed that all of the antenna ports are used in transmitting the signal, either in a coherent or non-coherent fashion. The alternative is to use the closest antenna port to the user for transmitting the signal. This way we minimize the required transmit power that also results in reducing the interference caused to other cells. As we will show in the numerical results, for a fixed average ergodic rate target, the single transmission mode increases the power efficiency of distributed antenna system significantly.

\section{Numerical Results}\label{sec:sim}
In this section we illustrate the performance of DAS and different algorithms we introduced in Sections \ref{sec:sol} and \ref{sec:ps}. We start with the placement optimization algorithms of Section \ref{sec:sol} and continue with different numerical studies for illustrating the power efficiency and area spectral efficiency of distributed antenna systems, under different assumptions on availability of CSI at the transmitter. 

\subsection{Convergence of the algorithm}
\begin{figure}
\begin{center}
\includegraphics[width=\figwwww]{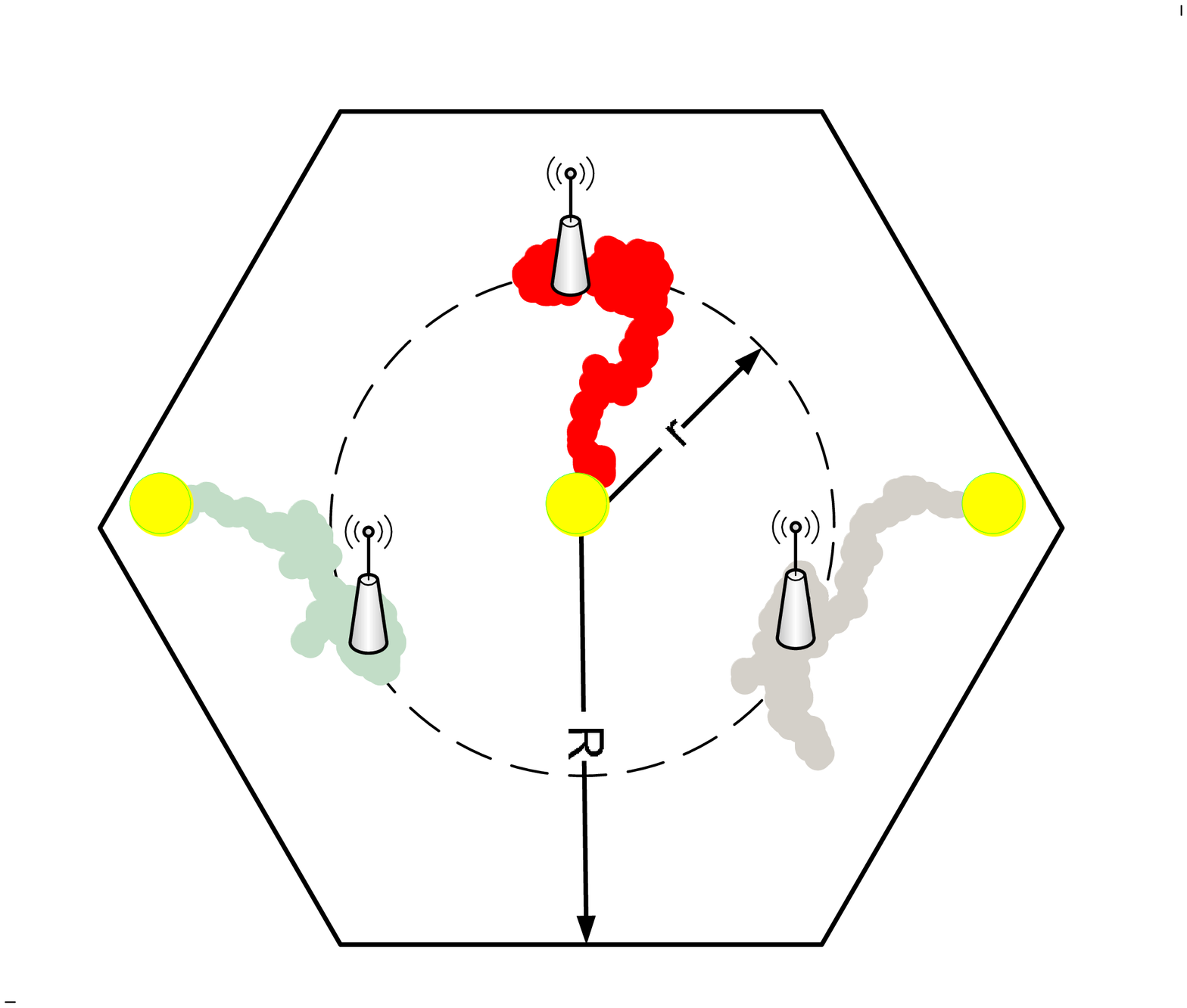}
\end{center}
\caption{Trajectory of the location of the ports in iterative algorithm with random initial points. Yellow points are the initial locations we started with}\label{fig:lm}
\end{figure}

 In this part, we mainly focus on the convergence of the stochastic update method. We consider a distributed antenna system with $N = 3$ ports, each equipped with a single antenna. We assume that the coverage region $\Pi$ of these three ports is a hexagon of radius $R = 1000$.  We assume independent log-normal shadowing of variance $\sigma_{sh} = 8$ between each port and the user. The path loss exponent is assumed to be $\alpha = 6$. We initialized the location of the antenna ports randomly inside the cell coverage region and the initial points are marked with yellow circles in Figure \ref{fig:lm}. Then we generated realizations of the random shadowing vector $\textbf{f}_t$ together with the random location $u_t$ uniformly inside the hexagonal cell region. We applied the algorithm in Section \ref{sec:sap} and plotted the trajectory of the locations of three ports as the algorithm evolved in Figure \ref{fig:lm}. The algorithm ran for 200k cycles. As we can see the optimum placement is symmetric with respect to the center of the cell region and they are all located in a hexagon of radius $r_{opt} = 550$. This radius is checked to be independent of the initial placement of the ports by many repetitions of the algorithm with different initial values.

\subsection{Comparison of different optimization approaches}
\begin{figure}
 \begin{center}
  \includegraphics[width=\figwwww]{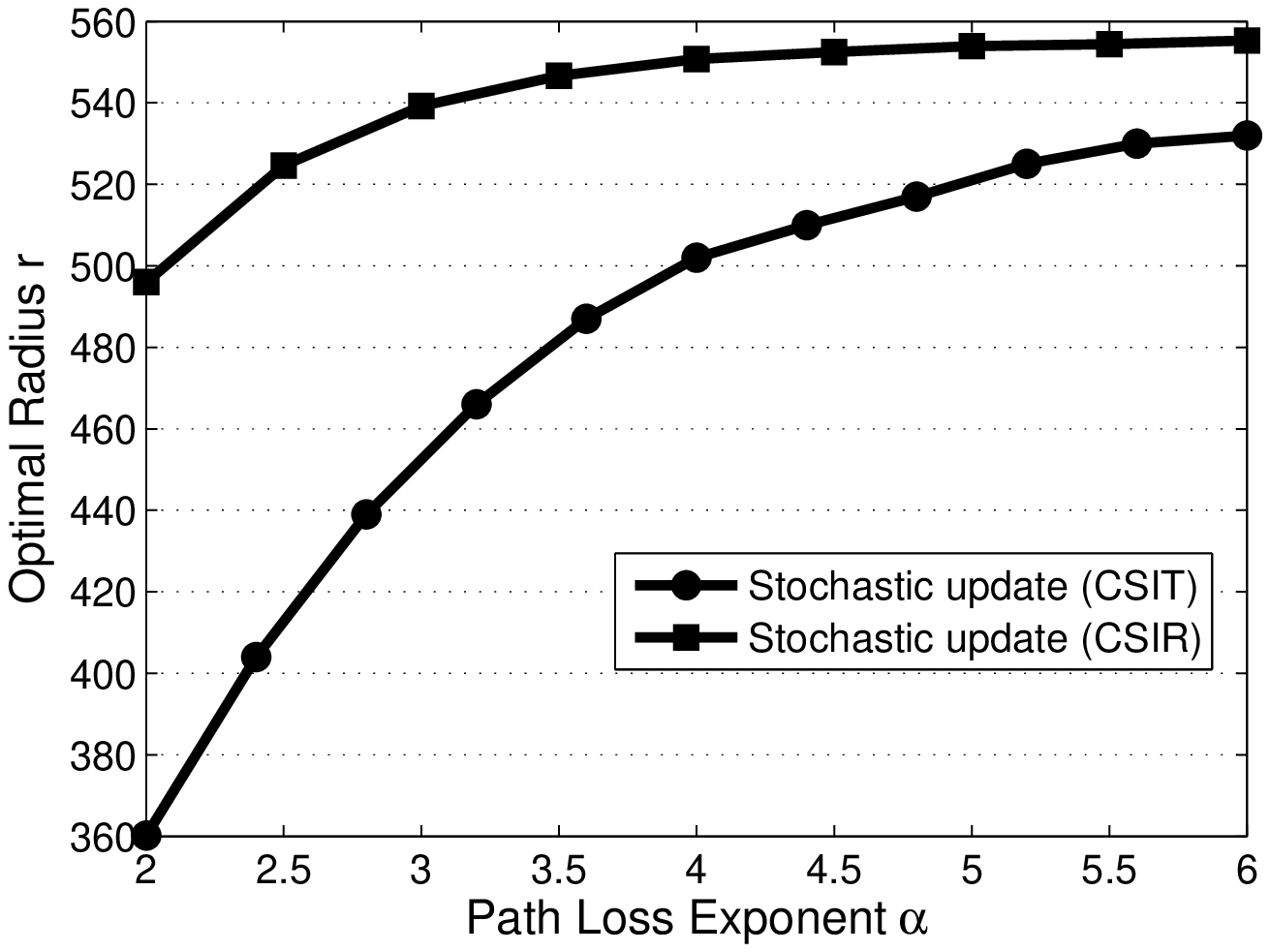}
  \end{center}
  \caption{Optimal placement radius $r$ of design example \ref{ex:1} for lower bound approximation approach vs stochastic update method in hexagonal cell of radius $R = 1000$  }\label{fig:ns}
\end{figure}

\begin{figure}		
 \begin{center}
  \includegraphics[width=\figwwww]{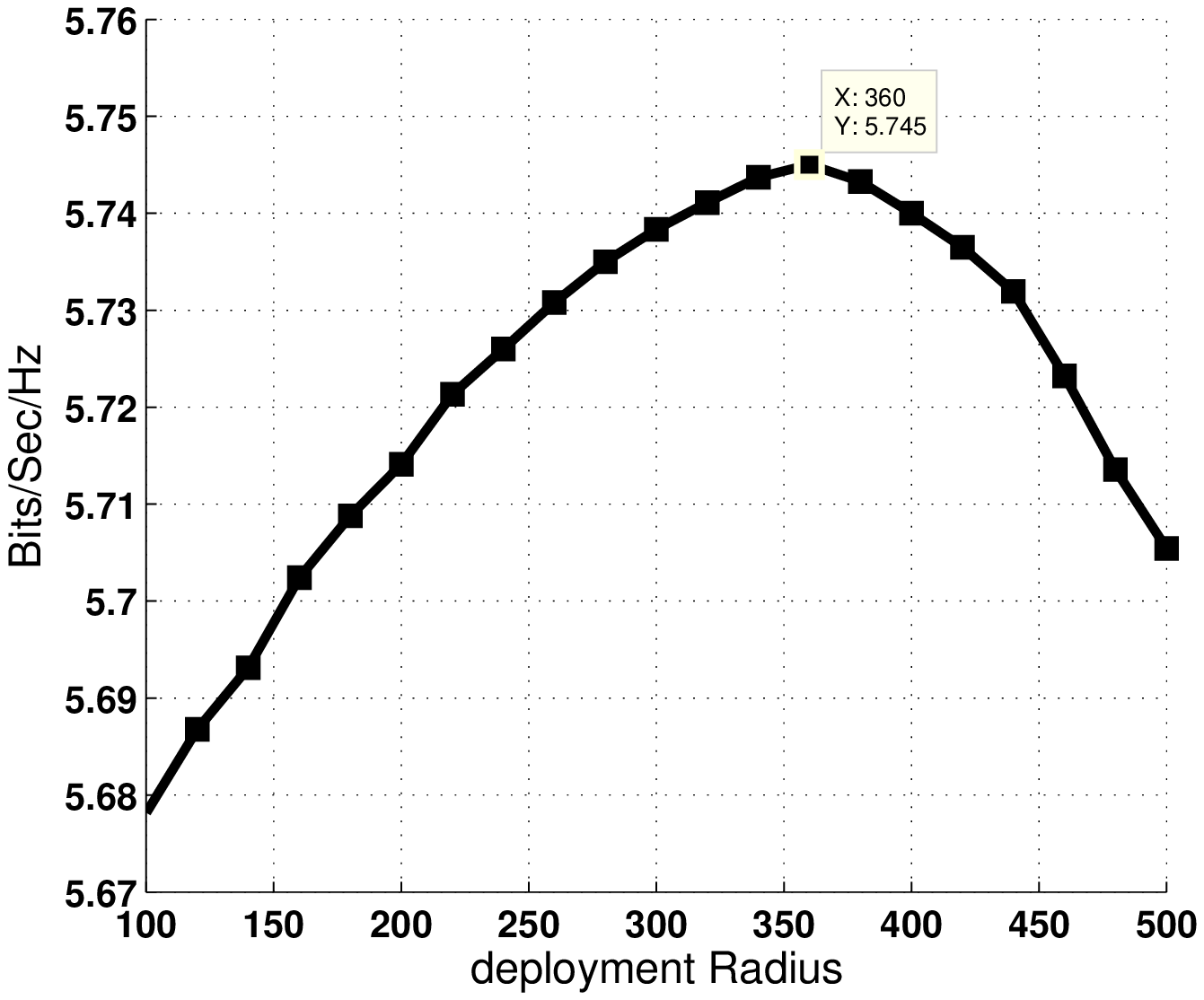}
  \end{center}
  \caption{Average ergodic rate of a single user in example \ref{ex:1} for deployment of antennas in different radius $r$ }\label{fig:ropt}
\end{figure}

In this part we compare the performance of the methods proposed in Sections \ref{sec:na} and \ref{sec:sap}. Figure \ref{fig:ns} illustrates the optimal radius $r$ to deploy three ports, obtained from two different approaches introduced in Section \ref{sec:sol}. The plot illustrates the optimal placement for different values of path loss $\alpha$. We used the stochastic update method for two different scenarios described in Section \ref{sec:cap} and plotted the optimal deployment radius as a function of path loss. As we can see, depending on using (\ref{bf}) or (\ref{ic}) for capacity, the optimal placement strategy is different, especially for small path loss exponent $\alpha$. In other words, different assumptions in Section \ref{sec:cap} for the cooperation among the ports and also availability of CSIT at the transmitter will change the optimal placement strategy. Note that these assumptions do not affect the lower bound. As we can see, for larger values of $\alpha$, the lower bound approximation approach creates a better estimate of the optimal deployment radius. This is understandable since the approximation we used in our lower bound optimization approach is tighter in high path loss regime. On the other hand, even for low path loss exponents, the stochastic update method converges to the optimal solution. In order to confirm this, for $\alpha = 2$, in Figure \ref{fig:ropt} we evaluated the expected ergodic rate of the system for different values of deployment radius $r$ with Monte-Carlo simulation, where we used (\ref{bf}) for capacity. As we can see, the optimal radius $r^* = 360$  is perfectly consistent with the optimal radius that the stochastic update method predicts. In general, as we can see in Figure \ref{fig:ns} the lower bound approximation approach gives a better estimation of the optimal placement when we assume that the distributed ports are transmitting independently.


\subsection{Power Efficiency}
\begin{figure}[ht]
\centering
\subfigure[Power gain without single transmission strategy]{
\includegraphics[width=\figwwww]{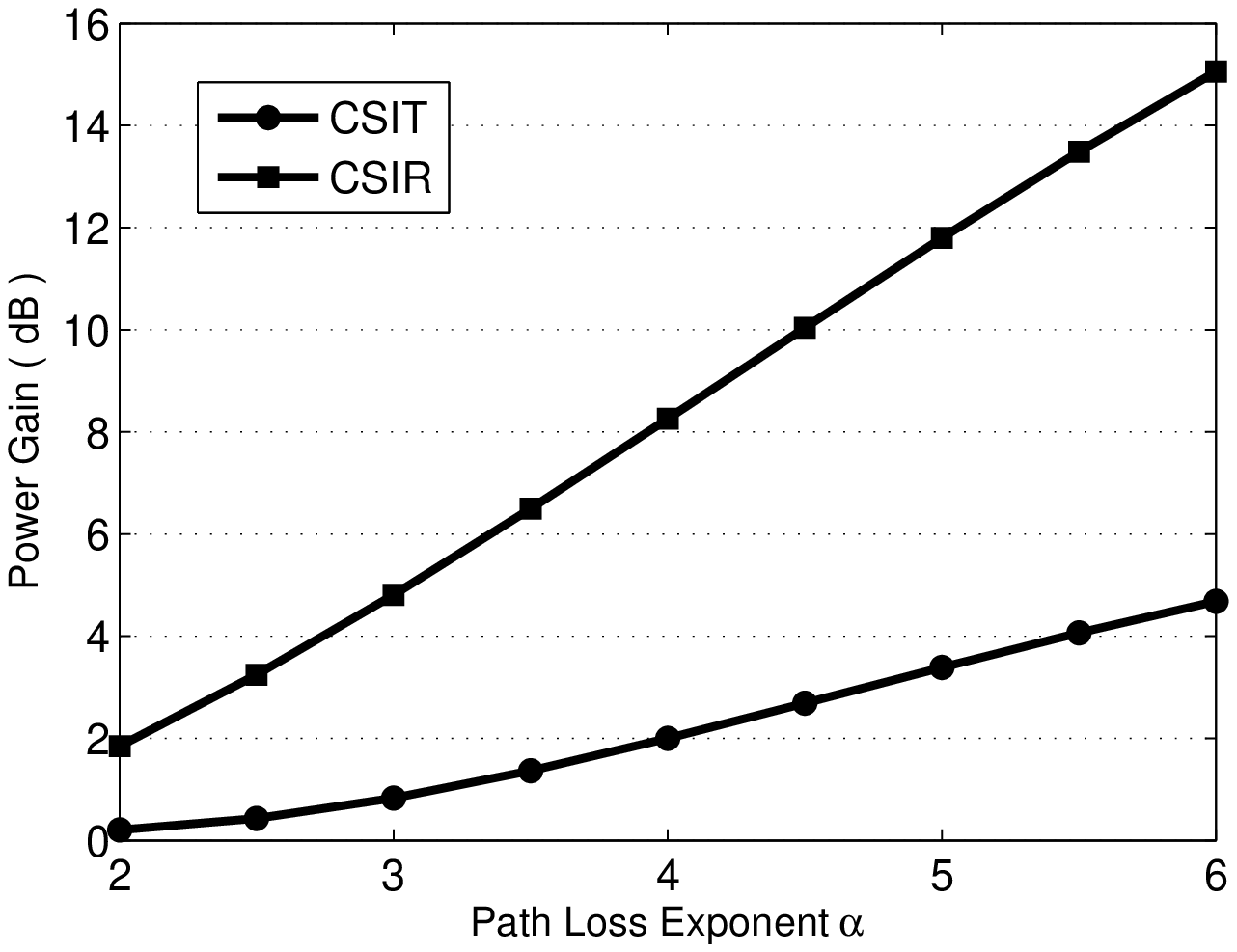}
\label{fig:pg}
}
\subfigure[Power gain with single transmission strategy]{
\includegraphics[width=\figwwww]{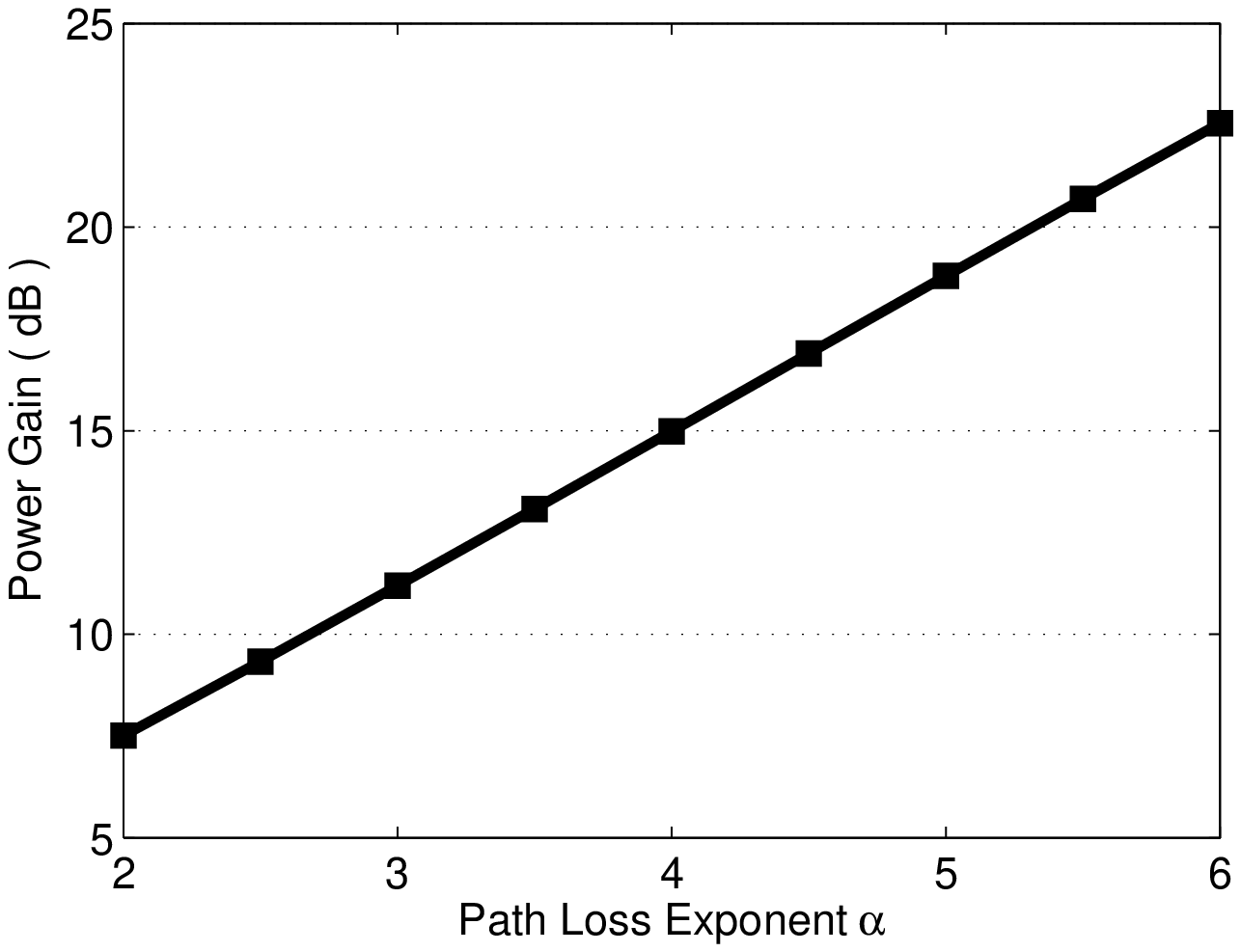}
\label{fig:pg2}
}
\label{fig:subfigureExample}
\caption[Optional caption for list of figures]{Power gain obtained from optimal placement of ports in design example \ref{ex:1}}
\end{figure}

Optimal placement will increase the power efficiency of the networks. In this part we aim to quantify this amount for $DAS(6,1)$, in comparison with the case where we have all of the antenna ports colocated in the center. In other words, for the case where we have the CSI only at the receiver and all antenna ports are co-located in the center of the cell, we set the per-antenna transmit power $S$ so that the average SNR at the edge of the coverage region is 10 $dB$. According to this transmit power for each value of $\alpha$, we calculate the average ergodic rate $C_t(\alpha)$ when all of the six antennas are co-located at the center. Then for each value of $\alpha$, we solved Problem \ref{problem1} in order to find the optimal placement of the ports inside the coverage region and also the minimum power required to achieve the same average ergodic rate $C_t(\alpha)$. Figure \ref{fig:pg} illustrates the ratio of the minimum powers required to satisfy $C_t$ in two cases. This gain is only due to the location optimization and as the figure indicates, the power gain increases with increased $\alpha$. The same procedure can be repeated for the case where the CSI is available at the transmitter as well. Figure \ref{fig:pg} also shows the gain that we get from optimally placing the ports when we have CSIT. As we can see, the power gain is considerably smaller when channel side information is available the transmitter. In other words, optimally distributing the antenna ports in the case of having CSIT doesn't help much, whereas the gain is noticeable when channel side information is not available at the transmitter, but only at the receiver. Further power saving is possible by utilizing the single transmission strategy of Section \ref{sts}. Figure \ref{fig:pg2} shows the power gain we achieve from using the single transmission scheme in $DAS(6,1)$ in comparison with the case where all of the six antenna ports are co-located in the center of the cell and we only have CSIR. Note that in the case of having CSIT, co-located ports can use a coherent transmission and reduce their transmit powers and yet achieve the same average ergodic rate. Therefore, the power gain from using single transmission strategy would be around $7.8$ $dB$ less, when the channel side information is available at the transmitter.

\subsection{Optimal layout is not always circular}
\begin{figure}[ht]
\centering
\subfigure[$N = 6$]{
\includegraphics[width=\figwwww]{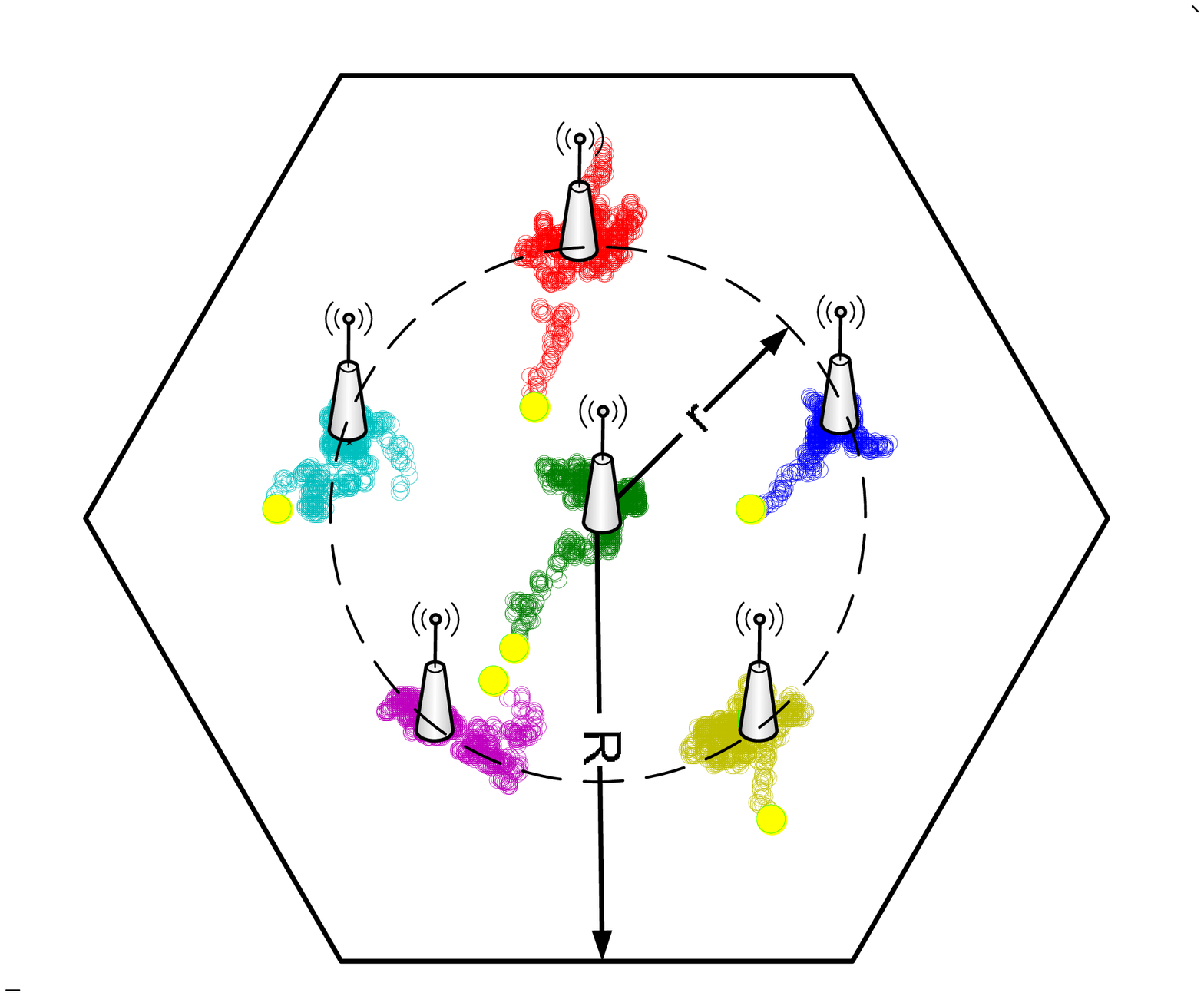}
\label{fig:op6}
}
\subfigure[$N = 12$]{
\includegraphics[width=\figwwww]{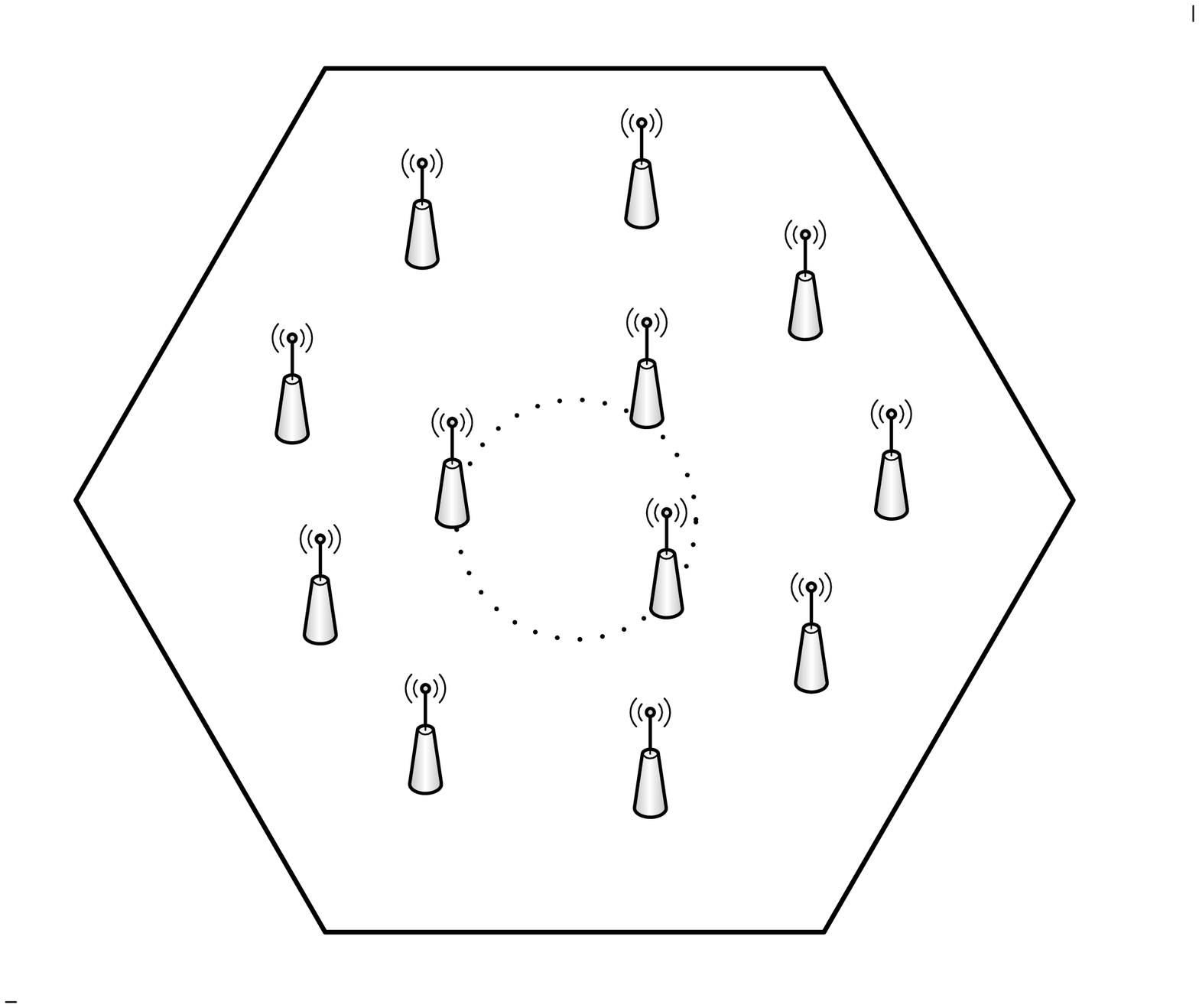}
\label{fig:op12}
}
\label{fig:op}
\caption[Optional caption for list of figures]{Optimal placement of N ports in an environment with path loss $\alpha = 5$.}
\end{figure}

In this part we run the stochastic update method in order to show that placing the antenna ports in a circular layout is not always optimal. In particular we run the stochastic update method to find the optimal placement of DAS $(6,1)$ in an environment with path loss $\alpha = 5$. Figure \ref{fig:op6} depicts the trajectory of the location of the ports through the iterations of stochastic update method, with random initial points. Yellow points in this figure are the initial locations we started with. As we can see the optimal deployment is not circular and one port should be located in the center. Figure \ref{fig:op12} also shows the optimal layout for the case where we have $N = 12$ ports. As we can see, in this case, the optimal layout has no port in the center of the cell.

\subsection{Interference effects}
\begin{figure}	
\begin{center}
\includegraphics[width=\figwwww]{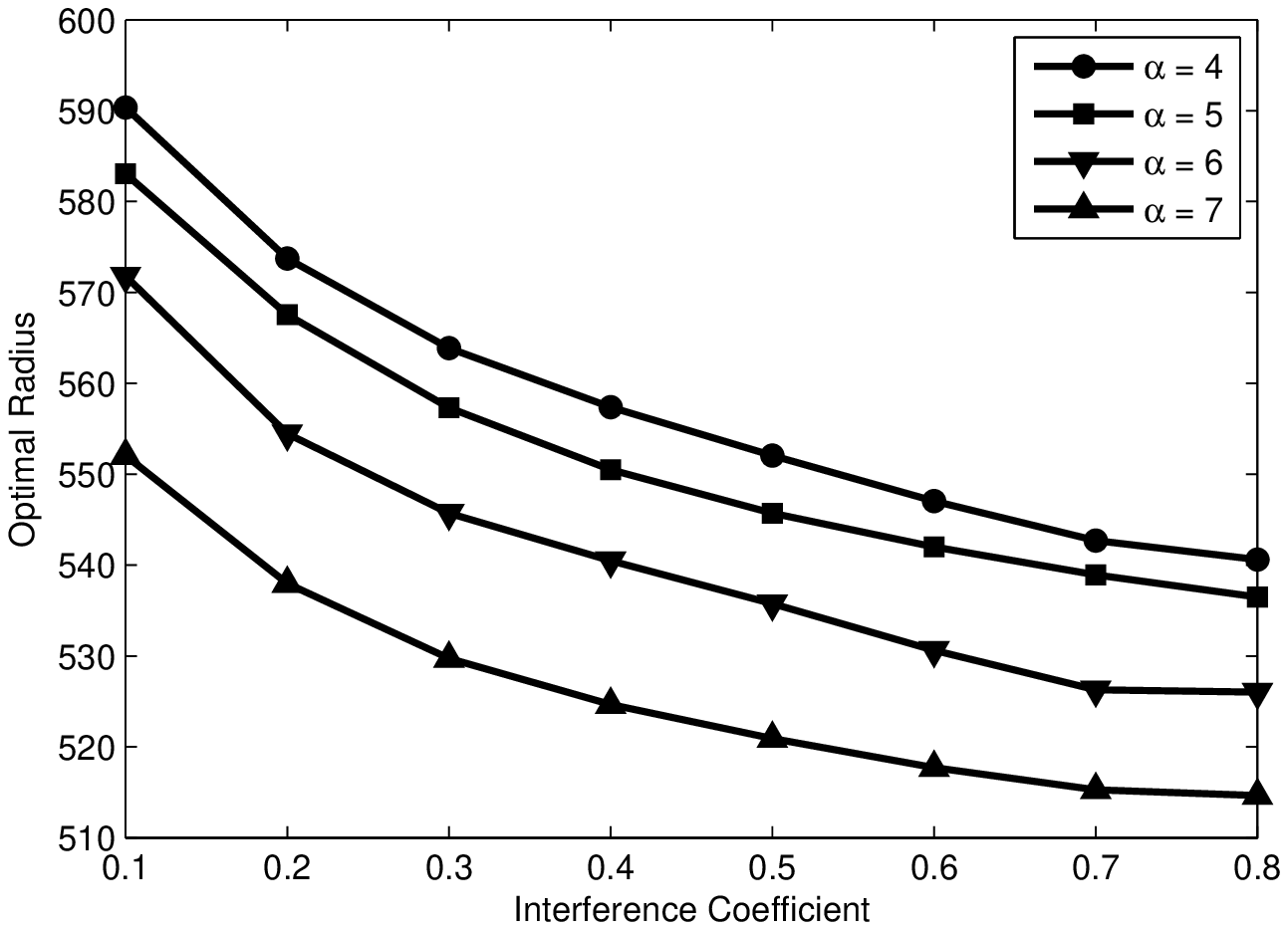}
\end{center}
\caption{Optimal radius $r$ of the antenna ports as a function of $\gamma$}\label{fig:IC}
\end{figure}

\begin{figure}	
\centering
\begin{center}
\includegraphics[width=\figwwww]{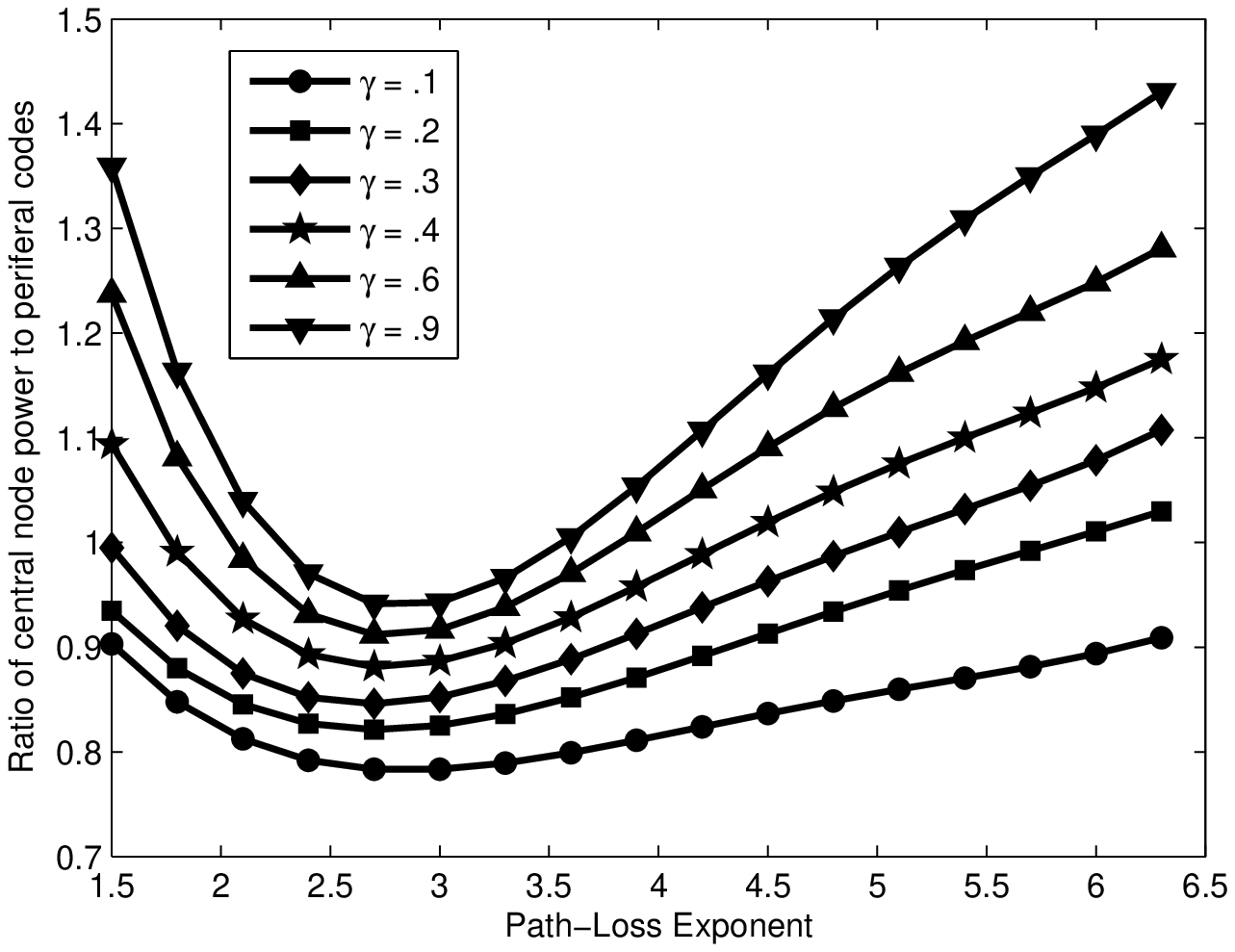}
\end{center}
\caption{Optimal positioning of the ports vs. path-loss exponent for different interference coefficient $\gamma$ s}
\label{fig:po}
\vspace{-0.3cm}
\end{figure}

In this part we study the effect of interference in the optimal placement of antenna ports, as described in Section \ref{sec:int}. For $N=7, L=1$,  we observe that the optimal layout always has one port in the center and six other ports in a circle of radius $r$ around it (see figure \ref{fig:In}), where $r$ depends on the interference coefficients $\gamma_j$ as well as the path-loss coefficient $\alpha$. In this part we set all the interference coefficients from neighboring cells $j = 1,2,...,6$ to $\gamma_j = \gamma$. We also let the stochastic gradient method find the optimal power allocation vector $\bar{S}^*$ for antenna ports. Figure \ref{fig:IC} plots the optimal radius $r$ of the antenna ports around the central port, as a function of interference coefficient $\gamma$, for different values of path-loss $\alpha$. As we see in the figure, the optimal layout shrinks towards the center of the cell as the interference coefficient $\gamma$ increases. For the power allocation, we observe that the optimal power allocation of all ports around the central one are the same. Figure \ref{fig:po} illustrates the ratio of the optimal power of the central port to the optimal power of the peripheral ports, as a function of $\alpha$.

\subsection{Cell size}
Studying the effect of cell size on our antenna port placement optimization framework is also of interest. If we fix the transmit power of the antenna ports, scaling down the cell size by a factor of $K$ leads to an increase of the overall power consumption of the network proportional to $K^2$. Therefore in order to have a fair comparison in this part, while scaling down the cell size, we also reduce the transmit powers by a factor of ${K^2}$. On the other hand, if we neglect the near field effect of $r_0$ in (\ref{fas}), scaling down the distances reduces the path-loss power falloff proportional to $K^{-\alpha}$. This leads to an increase of both the received powers from the antenna ports and interferers, by a factor of $K^{\alpha-2}$. Equivalently, we can say that scaling down the cell size by a factor of $K$ reduces the noise variance $\sigma_n^2$ in (\ref{int_noise}) proportional to $K^{\alpha-2}$. Hence for the path loss exponents $\alpha > 2$ and for large values of $K$, we will have an interference-limited system.

\begin{figure}
\centering	
\begin{center}
\includegraphics[width=\figwwww]{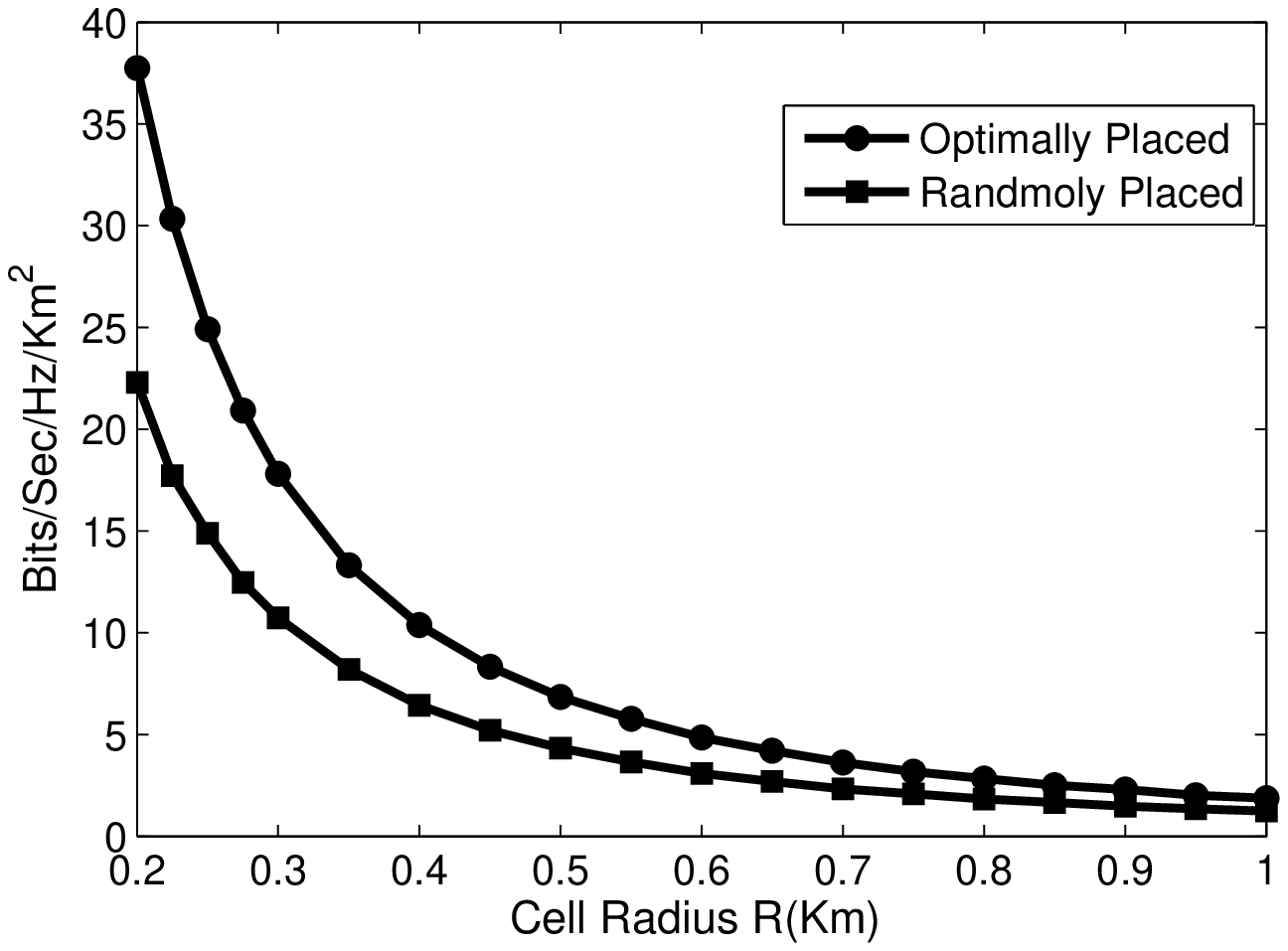}
\end{center}
\caption{Area spectral efficiency for a cell of radius $R$ for path loss exponent $\alpha = 3$, and interference coefficient $\gamma = 0.5$}
\label{fig:ase}
\vspace{-0.3cm}
\end{figure}

In order to illustrate the performance of optimally placed DAS as the cell size shrinks, we use the area spectral efficiency (ASE) metric in this part. This efficiency metric \cite{ase_} is defined as the average ergodic rate of the user per unit bandwidth per unit area, i.e
\begin{equation}
A(\ \textbf{P}, \bar{S}) = \frac{\bar{C}(\ \textbf{P}, \bar{S})}{\pi R^2},
\end{equation}
where $\textbf{P}$ and $\bar{S}$ are placement and power allocation of DAS in a cell of radius $R$. Figure \ref{fig:ase} illustrates this metric for $N=3$ single antenna ports, optimally placed in a cell with a varying radius\footnote{Note that for $N=3$ optimal power allocation is the same for all three ports}. The same figure also plots the area spectral efficiency of a system with $N=3$ ports, randomly placed inside the cell. For both cases we scale the power proportional to the area of the cell in order to have the same power consumption per unit area. Also for both cases we assumed that the CSI is only available to the receiver. As we see in the figure, in comparison with random placement, the optimal placement of the ports in DAS has a significant effect on area spectral efficiency of the system when we have CSIR. Adding this result with the fact that even randomly placed DAS outperforms colocated antenna systems \cite{ase}, we can conclude that optimally placed distributed antenna system can improve the spectral efficiency of the system significantly.

\section{Conclusion}\label{sec:con}
We established the downlink capacity of a single antenna receiver for generalized distributed antenna systems where each port has multiple transmit antennas.  We considered different assumptions about the availability of the channel states at the transmitter together with a per-port power constraint. For the case where CSI is available both to the transmitter and the receiver, we showed that  the optimal signaling for each of the antenna ports is beamforming, with the optimal beam weights proportional to the channel vector from the port to the receiver. For the case where the CSI is only available to the receiver but not to the transmitter, we showed that under a symmetric fading assumption, the optimal strategy for each port is to send independent signals with equal power from it's antennas. We have also presented a general framework for the placement optimization problem of distributed antennas in cellular networks. We concentrated on fully cooperative networks in which the base stations are connected together via an ideal backbone. We presented a fairly general framework with no implicit constraint on the antenna locations. As we showed in Section \ref{sec:int}, our stochastic optimization framework is sufficiently general to incorporate interference as well as general performance metrics. The proposed algorithm also works for arbitrary coverage regions. Since we assumed that we only have one tier of interferers in our SINR calculations, an obvious extension would be to incorporate the impact of frequency reuse on optimal placement of the ports. Finally we should mention that our optimization framework can also be used in the optimization of relay placement in cellular networks.

\appendices

\section{Capacity with CSIR}\label{app:capcsir}
In this appendix, we first generalize the proof of \cite[App. B]{mvu} in order to show that the optimal covariance matrix $Q^*$ that solves (\ref{rateg}) for a generalized $DAS(N,L)$, should be diagonal. Then we prove that the power budget in each antenna port should be equally allocated among the antennas. We denote the submatrices of the $Q$, the correlation of the Gaussian transmitted signals of the $m$th and $n$th ports, as $q_{mn} = \mathbb{E}\left[ x_m^H  x_n \right]_{L\times L}$\footnote{We dropped the index $0$ of the cell from the channel gain vector and transmitted signals }. First, note that $\textbf{tr}\left(q_{nn}\right)$ must be equal to $S_n$, otherwise, we can increase the diagonal elements of $q_{nn}$ until satisfying the constraint $\textbf{tr}\left(q_{nn}\right) \leq S_n$, and hence increase the objective. The main issue is to find the correlation matrices $q_{mn}$ for $m \neq n$.

The objective function in optimization problem (\ref{rateg}) can be expressed as
\begin{align}
C &= \mathbb{E}_{f} \left[ \log_2 \left( 1 + \frac{1}{\sigma_z^2} ~   \textbf{h} ~ \mathbb{E}\left[ x^H  x \right]  ~ \textbf{h}^H ) \right) \right]  \notag \\
&= \mathbb{E}_{f} \left[ \log_2 \left( 1 + \frac{1}{\sigma_z^2}  \sum_{m,n=1}^N  h_m \mathbb{E} \left[ x_m^H  x_n \right] h_n^H         \right) \right],\\
&= \mathbb{E}_{f} \left[ \log_2 \left( 1 + \frac{1}{\sigma_z^2}  \sum_{m,n=1}^N   h_m q_{mn} h_n^H          \right) \right],\\
&= \mathbb{E}_{f} \left[ \log_2 \left( 1 + \frac{1}{\sigma_z^2}  \sum_{n=1}^N  h_n q_{nn} h_n^H    +  \frac{1}{\sigma_z^2}  \sum_{m \neq n} h_m q_{mn} h_n^H         \right) \right].
\end{align}
Now, the key observation is that by exchanging $h_1$ by $-h_1$ the optimization problem does not change due to the symmetric distribution of $h_1$. Therefore, by flipping the sign of $h_1$, the objective function does not change 
\begin{align}
C &= \mathbb{E}_{f} \left[ \log_2 \left( 1 + \frac{1}{\sigma_z^2} h_1 q_{11} h_1^H + \frac{1}{\sigma_z^2}   \sum_{n=2}^N  h_n q_{nn} h_n^H  + \frac{1}{\sigma_z^2} \sum_{K_1}  h_m q_{mn} h_n^H   + \sum_{\bar{K_1}}  h_m q_{mn} h_n^H      \right) \right], \label{51} \\
&= \mathbb{E}_{f} \left[ \log_2 \left( 1 + \frac{1}{\sigma_z^2} h_1 q_{11} h_1^H + \frac{1}{\sigma_z^2}   \sum_{n=2}^N  h_n q_{nn} h_n^H  +  \frac{1}{\sigma_z^2} \sum_{K_1}  h_m q_{mn} h_n^H   - \sum_{\bar{K_1}}  h_m q_{mn} h_n^H      \right) \right], \label{52}
\end{align}
where $K_1 \overset{def}{=} \{ (m,n) | 2 \leq m,n \leq N , m \neq n  \} $, and $\bar{K}_1 \overset{def}{=} \{ (m,n) | 1 \leq m,n \leq N , m \neq n  \} \backslash K_1 $. Note that in (\ref{52}) we flipped the sign of $h_1$ and since either $m$ or $n$ equals 1 in the set $\bar{K}_1$, we should  flip the sign of the summation over $\bar{K}_1$. Using (\ref{51}) and (\ref{52}), we can write the objective function as
\begin{align}
C = \frac{1}{2} &\mathbb{E}_{f} \left[ \log_2 \left( 1 + \frac{1}{\sigma_z^2} h_1 q_{11} h_1^H + \frac{1}{\sigma_z^2}   \sum_{n=2}^N  h_n q_{nn} h_n^H  + \frac{1}{\sigma_z^2} \sum_{K_1}  h_m q_{mn} h_n^H   + \sum_{\bar{K_1}}  h_m q_{mn} h_n^H      \right) \right. \notag \\ 
&   \left.    \times  \left( 1 + \frac{1}{\sigma_z^2} h_1 q_{11} h_1^H + \frac{1}{\sigma_z^2}   \sum_{n=2}^N  h_n q_{nn} h_n^H  +  \frac{1}{\sigma_z^2} \sum_{K_1}  h_m q_{mn} h_n^H   - \sum_{\bar{K_1}}  h_m q_{mn} h_n^H      \right) \right] \\
&\leq \mathbb{E}_{f} \left[ \log_2 \left( 1 + \frac{1}{\sigma_z^2} h_1 q_{11} h_1^H + \frac{1}{\sigma_z^2} \sum_{n=2}^N   h_n q_{nn} h_n^H  +  \frac{1}{\sigma_z^2}  \sum_{K_1}  h_m q_{mn} h_n^H \right) \right] \label{55}
\end{align}
where equality in (\ref{55}) holds iff $\frac{1}{\sigma_z^2} \sum_{\bar{K_1}}  h_m q_{mn} h_n^H = 0$ for all $h_m, h_n$, where$ (m,n) \in \bar{K}_1$. In other words, the equality holds if and only if $ {q}_{1n} = {q}_{n1} = 0$ for all $n \neq 1$. Repeating the same procedure by flipping the sign for each of the channel vectors $h_m$ we can get the following inequality:
\begin{equation}
C \leq  \mathbb{E}_{f} \left[ \log_2 \left( 1 + \frac{1}{\sigma_z^2} \sum_{n=1}^N  h_n q_{nn} h_n^H \right) \right] ,
\end{equation}
where equality holds if $q_{mn} = 0$ for all $m \neq n$. Similarly, for each of the antenna ports, because of the independence and symmetry of all $NL$ channels in our model, we can repeat the same trick of flipping the sign of the channel gains for each antenna element to show that the off diagonal elements of the covariance matrix $q_{nn}$ should be zero, for all $1 \leq n \leq N$. In other words we can rewrite (\ref{rateg}) as  
\begin{align}
\text{Max} &:   \mathbb{E}_{f} \left[ \log_2 \left( 1 + \frac{1}{\sigma_z^2} \sum_{n=1}^N \sum_{i=1}^L  \| h_n(i) \|^2 q_{nn}(i,i)  \right) \right] \notag \\
\text{S.t} &: \sum_{i=1}^L   q_{nn}(i,i) \leq S_n , ~~~~~~~~~ \forall ~~ 1\leq n \leq N \label{10},
\end{align}
where $q_{nn}(i,i)$ is the $i$th diagonal element of $q_{nn}$. Now (\ref{10}) is a convex optimization problem with the optimum solution of $  q_{nn}(i,i) = S_n/L$ (can be obtained from KKT conditions). One way to show that the power budget in each port should be allocated equally between the antennas is to check the KKT conditions for optimality of (\ref{10}). The other way is to show that for any pair of antennas $i \neq j$ in the $n^{th}$ port, replacing the power allocations $  q_{nn}(i,i) $  and $  q_{nn}(j,j) $ by their average power $  \bar{q}_{nn} = \frac{1}{2} \left( q_{nn}(i,i) + q_{nn}(j,j) \right) $, the objective function in (\ref{10}) increases. To see that, consider the simple case where $N=1$ and we only have two antennas, i.e. $L=2$. Assume that the optimal values of allocated power for the antenna ports that maximizes (\ref{10}) are $q^*_{11}(1,1)$ and $q^*_{11}(2,2)$. Now we can write the optimal value of (\ref{10}) as
\begin{align}
 &\mathbb{E}_{f} \left[ \log_2 \left( 1 + \frac{1}{\sigma_z^2}    \left( \| h(1) \|^2 q^*_{11}(1,1) + \| h(2) \|^2 q^*_{11}(2,2) \right)  \right) \right] \notag \\
 &=  \mathbb{E}_{f} \left[ \log_2 \left( 1 + \frac{1}{\sigma_z^2}    \left( \| h(1) \|^2 (\bar{q} + \Delta) +  \| h(2) \|^2 (\bar{q} - \Delta) \right)  \right) \right] \label{26} \\
 &=  \mathbb{E}_{f} \left[ \log_2 \left( 1 + \frac{1}{\sigma_z^2}    \left( \| h(1) \|^2 (\bar{q} - \Delta) +  \| h(2) \|^2 (\bar{q} + \Delta) \right)  \right) \right] \label{27} \\
 &=  \mathbb{E}_{f} \left[ \frac{1}{2} \log_2 \left( 1 + \frac{1}{\sigma_z^2}    \left( \| h(1) \|^2 (\bar{q} - \Delta) +  \| h(2) \|^2 (\bar{q} + \Delta) \right)  \right) \right. \notag \\ 
 &+ \left. \frac{1}{2} \log_2 \left( 1 + \frac{1}{\sigma_z^2}    \left( \| h(1) \|^2 (\bar{q} + \Delta) +  \| h(2) \|^2 (\bar{q} - \Delta) \right)  \right) \right] \notag \\
 &\leq \mathbb{E}_{f} \left[ \log_2 \left( 1 + \frac{1}{\sigma_z^2}    \left( \| h(1) \|^2 \bar{q}  +  \| h(2) \|^2 \bar{q}  \right)  \right) \right], \label{28}
\end{align}
where $\Delta \overset{def}{=} q^*_{11}(1,1) - \bar{q}  $, and in deriving (\ref{27}) from (\ref{26}), we used the fact that the channel statistics for both of the antennas in the port ($h(1)$ and $h(2)$) are identical and independent. Therefore by swapping the index of antennas, the optimal value of the objective function should not change. The last inequality indicates that the optimal value of (\ref{10}) is less than (\ref{28}), which is a contradiction unless we assume that $q^*_{11}(1,1) = q^*_{11}(2,2)$, or equivalently $\Delta = 0$. This proves that the power should be equally allocated to both antennas for the case of having a single port and $L=2$. The proof can be easily generalized for any number of ports and also $L > 2$.

\section{Capacity with CSIT}\label{app:capcsit}
Since logarithm is a monotonically increasing function of its argument, maximizing $\log(1+\textbf{h} ~ Q  ~ \textbf{h}^H)$ in (\ref{rateg2}), is equivalent to maximizing the $SNR = \textbf{h} ~ Q  ~ \textbf{h}^H$. Therefore, the objective function  that we want to maximize in this part can be written as
\begin{equation}
 \textbf{h} ~ Q  ~ \textbf{h}^H  = \textbf{h} ~ \mathbb{E}\left[ x^H  x \right]  ~ \textbf{h}^H = \sum_{m,n=1}^N  h_m \mathbb{E} \left[ x_m^H  x_n \right] h_n^H \label{30} .
\end{equation}
Now we want to prove that for all $1 \leq m \leq N$ and $1 \leq n \leq N$, the following inequality holds
\begin{align}
 \Big\|  h_m & \mathbb{E}  \left[ x_m^H  x_n \right] h_n^H  \Big\| = \Big\| \sum_{i,j = 1}^L  h_m(i) \mathbb{E} \left[ x_m(i)^*  x_n(j) \right]  h_n(j)^* \Big\| \notag  \\
&\leq   \sum_{i,j = 1}^L \| h_m(i) \| ~ \| \mathbb{E} \left[ x_m(i)^*  x_n(j) \right] \| ~ \| h_n(j)^* \| \notag  \\
&\leq   \sum_{i,j = 1}^L \| h_m(i) \| ~ \| \mathbb{E} \left[ x_m(i)^*  x_n(j) \right] \| ~ \| h_n(j)^* \| \notag  \\
&\leq   \sum_{i,j = 1}^L \| h_m(i) \| ~  \sqrt{ \mathbb{E} \left[ \| x_m(i) \|^2 \right]  \mathbb{E} \left[ \| x_n(j) \|^2 \right] } ~ \| h_n(j)^* \| \notag  \\
&= \left( \sum_{i=1}^L \| h_m(i) \| ~  \sqrt{ \mathbb{E} \left[ \| x_m(i) \|^2 \right] } \right) \times \left( \sum_{i=1}^L \| h_n(i) \| ~  \sqrt{ \mathbb{E} \left[ \| x_n(i) \|^2 \right] } \right) \label{70} ,
\end{align}
where for deriving the last inequality, we used the Jensen's inequality together with the covariance inequality as $ \| \mathbb{E} \left[ x_m(i)^*  x_n(j) \right] \|  \leq  \mathbb{E} \left[ \| x_m(i)^*  x_n(j) \| \right]   \leq \sqrt{ \mathbb{E} \left[ \| x_m(i) \|^2 \right]  \mathbb{E} \left[ \| x_n(j) \|^2 \right]  }$. Also, from the power constraint  $ \sum_{i=1}^L \mathbb{E} \left[ \| x_m(i) \|^2 \right] \leq S_m $, we can conclude that $\sum_{i=1}^L \| h_m(i) \| ~  \sqrt{ \mathbb{E} \left[ \| x_m(i) \|^2 \right] } \leq \| h_m \| \sqrt{S_m} $, for all $ 1  \leq m \leq N$. Replacing this in (\ref{70}) results in
\begin{equation}
\Big\|  h_m  \mathbb{E}  \left[ x_m^H  x_n \right] h_n^H  \Big\| \leq \| h_m \|  \| h_n \| \sqrt{S_m S_n}.
\end{equation}
Then, from (\ref{30}) we can write the following inequality for the objective
\begin{align}
\notag \Big\|~ \textbf{h}  ~ Q ~   \textbf{h}^H ~ \Big\| = \Big\|  \sum_{m,n=1}^N    h_m  &\mathbb{E}  \left[ x_m^H  x_n \right] h_n^H  ~ \Big\| \leq  \sum_{m,n=1}^N  \| h_m \| \| h_n \|   \sqrt{S_m S_n} \label{cov1} \\
&=  \left( \sum_{n=1}^N   \| h_n \|   \sqrt{ S_n} \right)^2,
\end{align}
Now, note that the equality here holds if we replace the covariance matrix with \begin{equation}
Q_{csit}^* = \left[ q_1, q_2, ..., q_N \right]^H \left[ q_1, q_2, ..., q_N \right],
\end{equation}
where $q_n \overset{def}{=} \frac{\textbf{h}_n^*(0)}{\| \textbf{h}_n(0) \|} \sqrt{S_n}, ~~ 1 \leq n \leq N$. Since $Q_{csit}^*$ also satisfies the per-port power constraint in (\ref{rateg2}), we conclude that $Q_{csit}^*$ is indeed the solution of (\ref{rateg2}).

\clearpage
\pagestyle{empty}
\bibliographystyle{IEEEtran}
\bibliography{IEEEabrv,gcbib}

\end{document}